\documentclass[11pt,a4paper]{article}

\pdfoutput=1 % if your are submitting a pdflatex (i.e. if you have
\usepackage{jcappub}

\graphicspath{{figs/}{diagrams/}}

\usepackage{amsfonts}       % AMS fonts
\usepackage{amsbsy}     % Bolds (vectors, tensors, etc.)
\usepackage{units}

\usepackage{slashed}
\usepackage{multirow}
\usepackage{hhline}
\usepackage{comment}
\usepackage{subfig}
\usepackage{float}

\newcommand{\ii}{\mathrm{i}}

\newcommand{\tb}{\tan\beta}

\newcommand{\neut}[1]{\widetilde{\chi}^0_{#1}}
\newcommand{\cha}[1]{\widetilde{\chi}^+_{#1}}

\newcommand{\mue}{\mu_{\mathrm{eff}}}

\newcommand{\hobs}{H_{\rm obs}}

\newcommand{\mzero}{m_0}
\newcommand{\mhalf}{m_{1/2}}
\newcommand{\azero}{a_0}

\newcommand{\lam}{\lambda}
\newcommand{\kap}{\kappa}
\newcommand{\alam}{A_\lambda}
\newcommand{\akap}{A_\kappa}

\newcommand*{\sub}[1]{_{\mathrm{#1}}}

\newcommand{\NMT}{{\sc NMSSMTools}}
\newcommand{\MO}{{\sc MicrOMEGAs}}

\newcommand{\HiB}{{\sc HiggsBounds}}

\newcommand{\SI}{{\sc SuperIso}}

\newcommand{\gev}{\,\mathrm{GeV}}
\newcommand{\tev}{\,\mathrm{TeV}}

\newcommand{\refeq}[1]{{eq.~(\ref{#1})}}

\newcommand{\be}{\begin{equation}}
\newcommand{\ee}{\end{equation}}
\newcommand{\nn}{\nonumber}

\newcommand*\BR{\ensuremath{\rm BR}}

\newcommand*\bxsgamma{\ensuremath{\overline{B}\rightarrow X_{s}\gamma}}

\newcommand*\brbxsgamma{\ensuremath{\BR\left(\bxsgamma\right)}}

\newcommand*\brbsmumu{\ensuremath{\BR\left(B_s\to\mu^+\mu^-\right)}}

\newcommand*{\brbutaunu}{\ensuremath{\BR\left(B_u \rightarrow \tau \nu\right)}}

\subheader{
\textrm{\begin{flushright}
APCTP-PRE2015-015 \\
\end{flushright}}}

\title{Prospects for higgsino-singlino dark matter detection at IceCube and PINGU}
\author[a]{R. Enberg,}
\emailAdd{rikard.enberg@physics.uu.se}

\author[a,b]{S. Munir,}
\emailAdd{s.munir@apctp.org}

\author[a]{C. P\'erez de los Heros}
\emailAdd{cph@physics.uu.se}

\author[a]{ and D. Werder}
\emailAdd{dominik.werder@physics.uu.se}

\affiliation[a]{Department of Physics and Astronomy, Uppsala University,\\ 
             Box 516, SE--751 20 Uppsala, Sweden}

\affiliation[b]{Asia Pacific Center for Theoretical Physics, San 31, Hyoja-dong, \\
  Nam-gu, Pohang 790-784, Republic of Korea}

\abstract{
We study neutralino dark matter (DM) with large singlino fractions in the next-to-minimal
supersymmetric Standard Model (NMSSM). We perform a detailed analysis of 
the parameter space regions of the model that give rise to such singlino-dominated neutralinos 
while satisfying the
constraints from Higgs boson searches at the Large Electron Positron
(LEP) collider and the Large Hadron Collider (LHC), as well as from $b$-physics experiments. 
We find that this DM can yield a thermal relic density consistent with
the Planck measurement in
mass regions where the lightest neutralino of the minimal
supersymmetric Standard Model (MSSM) generally cannot. This is particularly true
for lighter DM masses, either less than 10\,GeV or between $60 -100$\,GeV, and for
heavier DM masses, between $500 - 1000$\,GeV. 
We then analyse the prospects for indirect detection of such 
DM at the IceCube neutrino telescope, assuming the complete
86-string configuration including DeepCore. We also consider the 
added sensitivity to low-mass DM with the proposed PINGU
extension. We find that IceCube is sensitive to some regions of the
NMSSM parameter space containing singlino-dominated DM and that 
a subset of such model points are already ruled out by the IceCube one-year data. 
IceCube will also be sensitive to some parameter space regions that
will not be probed by the upcoming ton-scale direct detection experiments. 
Moreover, we find that PINGU will be sensitive to DM in the 10 GeV range.
}

\begin{document}

\maketitle

%-------------------------------------------------------------------------------------------------
\section{Introduction \label{sec:intro}}
%-------------------------------------------------------------------------------------------------

The nature of the dark matter of the Universe is one of the major unresolved
problems in physics. Supersymmetric models with unbroken R-parity 
provide a possible candidate for a weakly interacting massive particle 
(WIMP), the lightest neutralino, which has been studied extensively
(see e.g.\ \cite{Jungman:1995df,Bergstrom:2000pn,Bertone:2004pz,Hooper:2007qk}).
There are good phenomenological and theoretical reasons apart from
the DM problem to consider such supersymmetric (SUSY) extensions of the
Standard Model (SM), including amelioration of the hierarchy problem or the
fine-tuning of the Higgs boson mass and unification of the gauge
couplings. 

The most commonly studied SUSY model is the MSSM\
\cite{Nilles:1983ge,Haber:1984rc}, which exhibits all of the above
benefits. The price for these benefits is a common mass term (called the
$\mu$-term) for the two Higgs doublets of the MSSM. 
This leads to another fine-tuning problem, namely why the 
$\mu$ parameter should lie at the
electroweak (EW) scale, as is required by phenomenology, when
its natural value would be the Planck scale. This is known
as the $\mu$-problem. Furthermore, the mass of the lightest Higgs
boson in the MSSM at the tree level has an upper limit equal to the
$Z$ boson mass. To fulfil the experimental constraints, large
corrections to the Higgs mass from the stop (and top) loops are
needed, which require a rather large stop mass, leading to an
additional source of fine-tuning in the model. If we take fine-tuning 
arguments seriously, we may consider exploring alternatives to the MSSM.

One such alternative is the NMSSM\ \cite{Fayet:1974pd,Ellis:1988er,Durand:1988rg,Drees:1988fc}, which has been reviewed in Refs.\ 
\cite{Ellwanger:2009dp,Maniatis:2009re}. In the NMSSM, in addition to
the two Higgs doublets of the MSSM  there is a singlet Higgs field,
which is the scalar component of a chiral singlet superfield added to
the MSSM superpotential. The reason for introducing this additional
scalar is that an effective $\mu$-term is now dynamically generated, so
  that the fine-tuned parameter $\mu$ is no longer needed (and can be
  removed by imposing a discrete $Z_3$ symmetry). This new
scalar mixes with the other scalars from the two doublets, leading to
a Higgs sector with seven Higgs bosons, compared to the five present
in the MSSM. In addition, the
  fermion component of the singlet superfield, the singlino, mixes
  with the higgsinos and gauginos, the fermion components of the
  Higgs doublet and gauge superfields,
  respectively, resulting in a total of five neutralinos. 

The interaction strengths of the neutralinos with other SM and SUSY
particles are governed by their masses and compositions. In the MSSM, the
neutralinos are broadly classified as gaugino-like, higgsino-like and
mixed gaugino-higgsino states, depending on the relative sizes of
the soft SUSY-breaking gaugino mass parameters, $M_{1,2}$,
and the $\mu$ -parameter in the neutralino mass matrix. If
$M_{1,2}$ are much larger than $\mu$, two of the
neutralinos are predominantly higgsino-like and the other two
are gaugino-dominated, while a smaller difference results in larger
gaugino-higgsino mixing. Similarly, in the NMSSM, the fifth
neutralino is either completely singlino-dominated or can be an admixture of the
singlino and, chiefly, the higgsinos. This purely singlino or mixed
higgsino-singlino state can be the
lightest of the five neutralinos and hence the lightest
supersymmetric particle (LSP) after
diagonalisation of the mass matrix. If so, some interesting
possibilities arise in the context of DM phenomenology, 
which are rather distinct from the MSSM. In particular, such a
neutralino can give the correct DM relic abundance with a considerably smaller
mass\ \cite{Abel:1992ts,Kozaczuk:2013spa,Cao:2013mqa,Han:2014nba,Ellwanger:2014dfa}
than is generally possible for the lightest neutralino in the MSSM, given the
LHC exclusion limits on SUSY.

A light singlino-dominated neutralino LSP in the NMSSM has been the subject of several studies recently. It is an important candidate for the events near $\sim10$\,GeV\ \cite{Kozaczuk:2013spa,Cao:2013mqa} reported by the CDMS II Collaboration\ \cite{Agnese:2013rvf}. With a mass near 35\,GeV, such a neutralino can also explain\ \cite{Cheung:2014lqa,Cao:2014efa,Feng:2014vea,Bi:2015qva} the $\sim 1-4$\,GeV 
$\gamma$-ray excess from the galactic centre reported by the Fermi
Large Area Telescope (LAT)\ \cite{Hooper:2010mq,Hooper:2011ti} (for
most recent results from Fermi-LAT, see\
\cite{Ackermann:2015zua}). Generally, the existence of a light DM
candidate is also supported by some other direct detection experiments
such as DAMA/LIBRA\ \cite{Bernabei:2010mq}, CoGeNT\
\cite{Aalseth:2010vx,Aalseth:2011wp} and CRESST-II\
\cite{Angloher:2011uu}. Experiments such as XENON100\
\cite{Aprile:2012nq}, LUX\ \cite{Akerib:2013tjd} or PICO~\cite{Amole:2015lsj} have released
strong exclusion constraints on the interaction strength of a generic
DM candidate with a nucleon for a considerably wide range of its
mass. Signatures of a higgsino-singlino LSP that can be probed at the LHC have been studied in\ \cite{Das:2012rr,Ellwanger:2013rsa,Kim:2014noa,Ellwanger:2014hia,Han:2015zba}. 
Indirect detection prospects for neutralino DM in the NMSSM have been studied in 
\cite{Ferrer:2006hy,Demidov:2010rq}.

In this article we revisit the neutralino sector of the NMSSM,
focusing on the LSP with a significant singlino fraction, which can
give sufficient relic abundance in the Universe. Through scans of 
the NMSSM parameter space, which were subject to
the most important experimental constraints, we find its regions where
such neutralino solutions are realisable. We then analyse the
potential of the IceCube neutrino telescope\ \cite{Achterberg:2006md}
to discover or exclude such LSP solutions. The annihilation of
neutralinos captured in the centre of the Sun results in a neutrino
flux at Earth. These neutrinos interact with ice at the South
Pole, producing leptons which emit Cherenkov radiation that can be
detected by IceCube. We take into account the winter data from a
complete 86-string configuration of IceCube, i.e., including also the
DeepCore strings, in order to perform a signal-to-background analysis
of the good higgsino-singlino solutions obtained from our scans. The impact
of the proposed extension PINGU\ \cite{Aartsen:2014oha} on the
statistical significance for mainly low mass DM is also investigated in
detail. 

The article is organised as follows. In section \ref{sec:nmssm} we
discuss the NMSSM and its neutralino sector. In section \ref{sec:scans} we
provide details of the model's parameter space of interest, the
procedure adopted for its scanning, the constraints imposed during
these scans and the impact of these constraints. In section
\ref{sec:dmsun} we explain the neutrino flux from the
Sun due to the annihilation of DM there.
In section \ref{sec:icecube} we highlight some important features of
the IceCube neutrino
observatory. In section \ref{sec:sb} we present the details of the
reconstruction of neutrinos from the Sun and the
background for this signal at the IceCube. In section \ref{sec:stat} we discuss the
statistical treatment of the signal process and the sensitivity that
can be reached at the IceCube after a given period of data
accumulation. In section \ref{sec:summary} we summarise our findings.

%-------------------------------------------------------------------------------------------------
\section{The neutralino sector of the NMSSM \label{sec:nmssm}}
%-------------------------------------------------------------------------------------------------

The $Z_3$-symmetric CP-conserving NMSSM has no mass parameters in the
superpotential, 
and is therefore also referred to as the scale-invariant NMSSM. It differs
from the MSSM by the omission of the $\mu$-term from the superpotential 
and the introduction of a singlet chiral superfield $\widehat S$,
while the Yukawa terms remain unchanged. The superpotential
is written as
\begin{equation}
 W\sub{NMSSM} = W\sub{Yukawa} + \lambda \widehat S \widehat H_u \widehat H_d + \frac{\kappa}{3} \widehat S^3\,,
\end{equation}
 where $\widehat H_u$ and $\widehat H_d$ are the two Higgs doublet superfields
of the MSSM. The scalar potential of the NMSSM constitutes of the
$D$-terms, which are the same as in the MSSM since $S$ does not couple
to the gauge bosons, the $F$-terms, plus the soft SUSY-breaking terms for the Higgs sector, which are given by
\begin{equation}
 V\sub{soft} = m_{H_u}^2 \left|H_u\right|^2 + m_{H_d}^2 \left|H_d\right|^2 + m_S^2 \left|S\right|^2
+ \left[ \lam \alam S H_u H_d +\frac{1}{3}\kap \akap S^3 + \text{h.c.}\right].
\end{equation}
The NMSSM scalar potential contains the new parameters $\lam, \kap, \alam,
\akap$, and $m_S$ compared to the MSSM. The parameters
$m_{H_u}^2$, $m_{H_d}^2$ and $m_S^2$ are
related to the $Z$ boson mass, $m_Z$, $\tb \equiv v_u/v_d$, where
$v_u$ and $v_d$ are the vacuum expectation values (VEVs) of $H_u$ and
$H_d$, respectively, and $s$, the VEV of the singlet field,
through the minimisation conditions
of the scalar potential for EW symmetry breaking, and can therefore be
eliminated. Here, instead of $s$ we will use
the effective $\mu$-term, $\mue=\lambda s$, as a free parameter. 
The VEVs of the doublet Higgs fields also satisfy the relation $v_u^2+v_d^2=v^2 =
2m_W^2/g_2^2 = (174 \gev)^2$, where $m_W$ is the mass of the $W$ boson
and $g_2$ is the $SU(2)_L$ gauge coupling.

The free parameters of the MSSM Higgs sector include $m_A$, the mass
of the CP-odd scalar, $\tb$ and $\mu$. In the NMSSM
$m_A$, which is the $1\times 1$ entry of the pseudo-scalar mass
  matrix before diagonalisation, gets traded for $\alam$ and
  $\mu$ gets replaced by $\mue$, so that all in all we have six Higgs
  sector parameters $(\lam, \kap, \alam, \akap, \tb,
  \mue)$. There are thus three more parameters in the NMSSM than in
  the MSSM, all of which originate in the Higgs sector.

As $S$ is a complex field, there are two additional physical
Higgs bosons in the NMSSM compared to the MSSM. For a CP-conserving
Higgs sector (as is assumed here), we have three CP-even neutral states
$H_1, H_2, H_3$ and two CP-odd neutral states $A_1$ and $A_2$, where
we take the states to be ordered in mass, so that $H_1$ and $A_1$ are the
lightest scalar and pseudo-scalar states, respectively. 

The fermion component of $\widehat S$ is called the singlino,
$\widetilde S$, which
mixes with the gauginos, $\widetilde B^0$ and $\widetilde W_3^0$, and
the higgsinos, $\widetilde H_d^0$ and $\widetilde H_u^0$. 
There are therefore five neutralinos in the NMSSM. At leading order
the neutralino masses and mixings depend on the parameters of the 
neutralino mass matrix. If we introduce the vector $\widetilde\psi^0 = (-\ii\widetilde B^0, -\ii\widetilde W_3^0, \widetilde H_d^0, \widetilde H_u^0, \widetilde S$), the non-diagonal mass Lagrangian in the gauge eigenstate basis is given by
\be
{\cal L}_\text{mass} = -\dfrac{1}{2} (\widetilde\psi^0)^T {\cal M}_{\widetilde{\chi}^0} \, \widetilde\psi^0 + \text{h.c.}\,,
\ee
where ${\cal M}_{\widetilde{\chi}^0}$ is the symmetric matrix
\be
{\cal M}_{\widetilde{\chi}^0} =
\begin{pmatrix}
M_1 	& 0 	& -\frac{g_1 v_d}{\sqrt{2}} 	& \frac{g_1 v_u}{\sqrt{2}} 	& 0  	\\
 	& M_2 	& \frac{g_2 v_d}{\sqrt{2}} 	& -\frac{g_2 v_u}{\sqrt{2}} 	& 0  	\\
 	&  	& 0				& -\mue				& -\lambda v_u \\
 	&  	& 				& 0				& -\lambda v_d \\
 	& 	& 				& 				& 2\kappa s 
\end{pmatrix} ,
\label{eq:massmatrix}
\ee
with $g_1$ being the $U(1)_Y$ gauge coupling. 
 
The neutralino masses and compositions at the tree level thus depend
on the Higgs sector parameters $\lam,\kap,\mue,v_u,v_d$ and the
gaugino masses $M_1$ and $M_2$. The mass matrix in \refeq{eq:massmatrix}
can be diagonalised by a unitary matrix $N$ to give
$D=\text{diag}(m_{\neut{i}}) = N^* {\cal M}_{\widetilde{\chi}^0} N^\dag$,
for $i=1 - 5$. If all the parameters in the mass matrix are real (which
we assume to be the case here) then the matrix $N$ is orthogonal, and
we have $D= N {\cal M}_{\widetilde{\chi}^0} N^T$. The neutralino mass
eigenstates are then given by $\neut{i} = N_{ij} \widetilde\psi^0_j$.
 
The eigenvalues of $N_{ij} $ are all real, but can be positive or
negative. (They can be made positive by a phase transformation.) 
They are not ordered in mass after performing the diagonalisation, 
but should then be reordered so that $\neut{1}$ is the lightest 
neutralino. The linear combination,
 \be
 \neut{1} = N_{11} \widetilde B^0 + N_{12} \widetilde W_3^0 + N_{13} \widetilde H_d^0 + N_{14} \widetilde H_u^0 + N_{15} \widetilde S^0\,,
 \ee
is thus our DM candidate. In order to describe the composition of
$\neut{1}$, we define the gaugino fraction in it as
$Z_g=|N_{11}|^2+|N_{12}|^2$, the higgsino fraction as
$Z_h=|N_{13}|^2+|N_{14}|^2$ and the singlino fraction as
$Z_s=|N_{15}|^2$. As noted earlier, the focus of this study is a
$\neut{1}$ with a non-negligible singlino fraction, which we define to
be $Z_s \geq 0.05$. Such a $\neut{1}$ could lead to some distinct 
phenomenological scenarios precluded in the MSSM.

Let us note some properties of the tree-level neutralino mass matrix
given in \refeq{eq:massmatrix}, assuming the mass parameters $M_1$
and $M_2$ of the gauginos to be very heavy so that they are decoupled
from the $3\times 3$ higgsino-singlino block. The
$5\times 5$ diagonal element, $2\kap s = 2\kap \mue /\lam$, corresponds to
the mass of the singlino. Thus, if this is small the $\neut{1}$ is
more likely to be singlino-dominated after diagonalisation. Similarly, if
$\mue$ is small, depending on the size of $\lam$, the lightest
neutralino can instead have a larger higgsino component. To have a
singlino-dominated WIMP, we therefore need a large $\mue$ and a
small $\kap$. A large $\lam$ also reduces the
size of the $5\times 5$ term, but at the same time it enhances
the sizes of the off-diagonal terms, leading to a larger
mixing. Note, however, that the presence of a certain amount of higgsino in the WIMP
is necessary to obtain a realistic relic abundance. Thus 
$\mue$ should not be too large and $\lam$ should not be too small. 
Evidently, the composition of such a WIMP is quite insensitive to the
value of $\tb$. In reality the situation is more complex, as the other
elements of the matrix also need to be considered. Moreover, just as
 in the MSSM, this mass matrix is subject to radiative corrections\
\cite{Ellwanger:1993hn,Elliott:1993ex,Elliott:1993uc,Elliott:1993bs,Pandita:1993hx,Ellwanger:2005fh,Staub:2010ty,Ender:2011qh}
from various other sectors of the NMSSM. Thus, at higher orders the parameters
of these sectors also become crucial. 

\section{Model scans and constraints\label{sec:scans}}

As noted above, beyond tree level the parameters of other
SUSY sectors also need to be taken into account along with the Higgs
sector ones, when drawing inferences about the neutralino sector. 
However, the most general EW-scale NMSSM contains more than
 130 parameters in total. One way to reduce the number of free parameters is to
consider a minimal supergravity-like scenario in which all the
dimensional parameters are defined at the grand unification (GUT) scale and certain
universality conditions are imposed on them. In this so-called
constrained NMSSM (CNMSSM) (for some recent analyses, see, e.g.,\
\cite{Kowalska:2012gs,Kim:2013uxa,Fowlie:2014faa}) the scalar soft 
SUSY-breaking masses are
unified into a generic mass parameter $\mzero$, the gaugino masses into
$\mhalf$, and all the trilinear couplings into $\azero$. Then, given $m_Z$, these three
parameters along with the coupling $\lam$, taken as an input at the
SUSY-breaking scale, $M_{\rm SUSY}$, and the sign of $\mue$ constitute 
the only free parameters. All the parameters at $M_{\rm SUSY}$
 are then obtained from these GUT-scale parameters using the
renormalisation group equations. 

In order to allow more freedom in adjusting the Higgs and neutralino
sector parameters, one can relax the above-mentioned universality conditions partially.
In that case, similarly to the MSSM with non-universal Higgs masses (NUHM), 
$m_{H_u}$, $m_{H_d}$ and $m_S$
are disunified from $\mzero$ at the GUT
scale and taken as free parameters, which can then be traded 
for $\kap$, $\mue$ and
$\tb$ at the EW scale. Also, $a_\lambda$ and $a_\kappa$, the
GUT-scale input values, respectively, of the EW-scale parameters
$\alam$ and $\akap$, are disunified from $\azero$. This results in a total of nine
input parameters and we refer to this model as the next-to-NUHM (NNUHM) here.

Alternatively, without imposing the GUT-universality conditions, a convenient way
to considerably reduce the number of parameters in the general
EW-scale NMSSM is to assume that the matrices for the sfermion
masses and for the trilinear scalar couplings are
diagonal. Furthermore, the soft SUSY-breaking parameters are taken to
be real and those corresponding to the first two
generations are unified. Thus, with
only about 25 or so parameters in total, one can study the
most important low-energy characteristics of the model.
In our analysis, while requiring the $\neut 1$
to have $Z_s \geq 0.05$, we also want to take into account all of
 its possible annihilation and
co-annihilation channels resulting in the correct relic
abundance. Some of these co-annihilation channels require the existence
of a sfermion not much heavier than the $\neut 1$ itself. However, in
order that the model candidate for the SM-like Higgs boson, $\hobs$, 
observed at the LHC\ \cite{Aad:2012tfa,Chatrchyan:2012ufa,Chatrchyan:2013lba} 
has a mass near $125\gev$,
large corrections from the SUSY sector are necessary, implying
sufficiently large sfermion mass parameters. Crucially though, these
corrections are almost entirely dominated by the stops. 
In view of all these considerations, we impose the following
universality conditions on the parameters of the general NMSSM:
\begin{gather}
M_{\widetilde{Q}} \equiv M_{\widetilde{Q}_{1,2}}  = M_{\widetilde{U}_{1,2}}  = M_{\widetilde{D}_{1,2}}\,, \nn\\
M_{\widetilde{L}} \equiv M_{\widetilde{L}_{1,2,3}} = M_{\widetilde{E}_{1,2,3}}\,,  \nn\\
M_2 =  \frac{1}{3} M_3 \,, \nn\\
A_0 \equiv A_t = A_b = A_{\tau}\,, \nn
\end{gather}
where $M_{Q_{1,2}},\,M_{U_{1,2}},\,M_{D_{1,2}}$
 are the soft masses of the squarks from the first two 
generations, $M_{L_{1,2,3}}$ and $M_{E_{1,2,3}}$ the soft slepton
masses, $M_{1,2,3}$ the soft gaugino masses and $A_{t,b,\tau}$ the 
trilinear Higgs-sfermion couplings.\footnote{Note in 
particular that a unique soft mass has been used for all three
generations of the sleptons. This is not the case for the squarks, for which the 
soft masses of the first two generations are disunified from the
respective ones, $M_{\widetilde{Q}_3}$, $M_{\widetilde{U}_3}$ 
and $M_{\widetilde{D}_3}$, of the third generation.} 
This leaves us with a total of 14 free parameters to scan over. 
In the following we will refer to this model as the NMSSM-14.

In order to obtain maximum possible configurations of the free parameters
that yield the desired neutralino composition while satisfying
important constraints coming from various experimental sources, these
parameters need to be scanned numerically.
We first carried out scans for the NNUHM, using the
MultiNest-v2.18\ \cite{Feroz:2008xx} package, which is
linked to the public code \NMT-v4.2.1\ \cite{Ellwanger:2004xm,Ellwanger:2005dv,Das:2011dg,NMSSMTools}
to calculate the SUSY mass spectra and branching ratios (BRs) as well as the
Higgs boson signal rates. The signal rate is defined, for the $X$ decay
channel of a given NMSSM Higgs boson, $H_i$, as
 \begin{equation}
R_i^X \equiv  \frac{\sigma(gg\rightarrow H_i)\times {\rm BR}(H_i\rightarrow
  X)}{\sigma(gg \rightarrow  h_{\rm SM})\times {\rm BR}(h_{\rm SM} \rightarrow X)}\,,
\end{equation}
where $h_{\rm SM}$ is the SM Higgs boson with the same mass as $H_i$. 

We consider two distinct scenarios for the Higgs sector:\ one scenario
where the $\hobs$ corresponds to the lightest NMSSM Higgs boson,
$H_1$, and one where the $\hobs$ corresponds to the heavier $H_2$. In the latter scenario, there is thus a Higgs boson with mass less than 125\,GeV that remains to be discovered. Both of these scenarios are easily realisable
in significant portions of the NMSSM parameter space\ \cite{Ellwanger:2011aa,King:2012is,Cao:2012fz,Ellwanger:2012ke,King:2012tr,Gherghetta:2012gb}. 
Separate scans were performed for the NNUHM for these two scenarios.
A $H_i$, with $i=1,\,2$, is identified with the $\hobs$ by requiring it to have a
mass lying in the $122 - 128$\,GeV range\footnote{The
extended mass range is to allow up to 3\,GeV uncertainty in the
theoretical prediction of the mass of the assumed $\hobs$.} and
$R_{\rm obs}^X$ consistent with the
experimentally measured $\mu^X\equiv \sigma(pp\rightarrow \hobs
\rightarrow X)/ \sigma(pp\rightarrow h_{\rm SM} \rightarrow X)$ within $\pm 1\sigma$ error, for each $X$. 
This condition assumes that the inclusive $pp$ cross section for Higgs boson
production at the LHC can be approximated by the dominant gluon fusion 
channel. Note, however, that in the rare cases when
$\sigma(gg\rightarrow \hobs \rightarrow X)/\sigma(gg\rightarrow h_{\rm SM} \rightarrow X)$ for
a given $X$ is also provided by the experimental collaboration, we
use it instead as $\mu^X$ for comparing $R_{\rm obs}^X$ with. 

The most recent publicly released values of $\mu^X$ measured by the ATLAS and CMS
collaborations are given in table~\ref{tab:rates}. Instead of
considering the two sets of experimental measurements of the same 
observables separately in two scans for the given
$\hobs$ case, we delineate an `optimal allowed range' of each
$R_{\rm obs}^X$. This range spans
min[$\mu^X-|1\sigma|^{\vee}$\,(ATLAS), $\mu^X-|1\sigma|^{\vee}$\,(CMS)] to
max[$\mu^X+|1\sigma|^{\wedge}$\,(ATLAS), $\mu^X+|1\sigma|^{\wedge}$\,(CMS)], where
$\mu^X$ denotes the central value of the measurement and
$|1\sigma|^\wedge$ and $|1\sigma|^\vee$ imply, respectively, the positive and
negative error on it, as long as
$\mu^X-|1\sigma|^{\wedge}$ for one experiment is not higher than 
$\mu^X-|1\sigma|^{\vee}$ for the other. This range is also given
in the last column of table~\ref{tab:rates} for each $X$.
One can notice in the table that while $\mu^{\gamma\gamma}$ from
the two collaborations are mutually consistent, the ATLAS measurement
of $\mu^{ZZ}$ is considerably higher than the CMS one.  
Note also that since the $WW$ and $ZZ$ decays of a given $H_i$ depend on
the same $VVH_i$ reduced coupling, \NMT{} computes a unique value, $R_{\rm
  obs}^{VV}$, of the signal rates for both these
channels. Therefore, our defined optimal range of $R_{\rm  obs}^{VV}$ covers
$\mu^{ZZ}-|1\sigma|^{\vee}$ from CMS to
$\mu^{WW}+|1\sigma|^{\wedge}$ from ATLAS. We ignore 
the $\mu^{bb}$ measurement here since, aside from the fact that the
error on it is very large, it does not take into account the
gluon fusion mode for the production of $\hobs$.

\begin{table}[tbp]
\begin{center}
\begin{tabular}{|c|c|c|c|c|c|}
\hline
\multirow{2}{*}{$X$} &
\multirow{2}{*}{$\mu^X$(CMS)\ \cite{Khachatryan:2014jba}}
&\multirow{2}{*}{$\mu^X$(ATLAS)} & \multirow{2}{*}{Allowed $R_{\rm obs}^X$ range} & \multicolumn{2}{|c|}{Observed $R_{\rm obs}^X$ range} \\
&&&& $H_{\rm obs} = H_1$ & $H_{\rm obs} = H_2$ \\
\hline
\hline
$\gamma \gamma$ & $1.13 \pm 0.24$ & $1.17 \pm 0.27$\ \cite{Aad:2014eha}
& 0.89 -- 1.37 & 0.91 -- 1.1 & 0.89 -- 1.12 \\
\hline
$ZZ$ & $1.0 \pm 0.29$ &
$1.44^{+0.40}_{-0.35}$\ \cite{ATLAS-CONF-2014-009}
&\multirow{2}{*}{0.71 -- 1.31} &\multirow{2}{*}{0.95 -- 1.05} &\multirow{2}{*}{0.88 -- 1.05} \\
\hhline{---~~~}
$WW$ & $0.83 \pm 0.21$ & $1.09^{+0.23}_{-0.21}$\ \cite{ATLAS:2014aga} && &\\
\hline 
$\tau\tau$ & $0.91 \pm 0.28$ &
$1.4^{+0.5}_{-0.4}$\ \cite{ATLAS-CONF-2014-009} & 0.63 -- 1.9 & 0.9 --
1.01 & 0.63 -- 1.06\\
\hline
 \end{tabular}
\caption{Higgs boson signal rates measured by the CMS
  and ATLAS collaborations, their ranges enforced in the scans and
  those observed for the good points from the scans corresponding to the $\hobs = H_1$ and $\hobs
  = H_2$ cases.}
\label{tab:rates}
\end{center}
\end{table}

During our scans the $b$-physics observables were subject to the
following constraints.
\begin{itemize}
\item $2.63 \times 10^{-4} \leq \brbxsgamma \leq 4.23 \times 10^{-4}$\ \cite{Barberio:2008fa,Amhis:2014hma,heavyflavor},
\item $0.71 \times 10^{-4} < \brbutaunu < 2.57 \times 10^{-4}$\ \cite{Barberio:2008fa,Amhis:2014hma,heavyflavor},
\item  $1.3 \times 10^{-9} < \brbsmumu < 4.5 \times 10^{-9}$\ \cite{Aaij:2013aka,Chatrchyan:2013bka,CMSandLHCbCollaborations:2013pla}.
\end{itemize}
The theoretical evaluation of these observables was carried out using\ 
\SI-v3.4\ \cite{superiso} and the above ranges are
the ones allowed at the 95\% confidence level (CL), as suggested in the manual of
the package. Each SUSY point was also required to be consistent with the
LEP and LHC exclusion limits on the other, non-SM-like,
Higgs bosons of the model, as tested using \HiB-v4.1.3\
\cite{Bechtle:2008jh,Bechtle:2011sb,Bechtle:2013gu,Bechtle:2013wla}.
The satisfaction of the LEP limit on $\cha{1}$ mass and the
perturbativity constraints on the various Higgs boson couplings 
are ensured by \NMT\ intrinsically. While it also tests each point against
the squark/gluino and slepton searches from the LHC, for 
added surety we further required the scan to allow a 
point only if it gave a gluino mass larger than 1400\,GeV.
Finally, we enforced an upper limit on the $\neut{1}$ thermal relic
density, $\Omega_{\neut 1} h^2$, to be less than 0.131, thus allowing up to $+
10\%$ error in its theoretical calculation (and in this way taking
into account also the experimental error), performed by the 
package \MO-v4.1.5\ \cite{Belanger:2013oya},
given the measured value of 0.119 from the Planck
telescope\ \cite{Ade:2013zuv,Planck:2015xua}. No lower
limit is imposed on $\Omega_{\neut 1} h^2$ in view of the
possibility that $\neut{1}$ can make up 100\% of the DM in the universe
via, e.g., non-thermal production\ \cite{Profumo:2004at}, despite yielding an
under-abundant thermal relic density. There is also the alternative
possibility of the $\neut{1}$ constituting only a fraction of a multi-component DM
(see, e.g., \cite{Zurek:2008qg}). 

Our scans showed that the NNUHM serves as a good approximation for
the general NMSSM for the $\hobs = H_2$ scenario, since we
obtained a sufficiently large
number of points with the desired neutralino composition. We point out
here that in this scenario, the $H_2$
can obtain a large tree-level mass in a more `natural' way, i.e.,
without requiring very large corrections from the SUSY sectors and
hence forcing sparticle masses to be large, for its physical mass to lie near
125\,GeV. This happens for $\lam
\gtrsim 0.1$ and $\tb$ not too large. The scanned ranges 
of the input parameters were 
chosen in light of these observations, and are
listed in table~\ref{tab:params}(a). Also, in this scenario the $\neut 1$, typically
a higgsino-singlino mixture, is generally light but can still give the
correct $\Omega_{\neut 1} h^2$. The singlino and the higgsino masses are pushed
towards smaller values for two main reasons: a) $\lam$ is generally
large and b) the masses of singlet-like
$H_1$ and $A_1$, which scale with $\kap s$, are smaller than
 the $\hobs$ mass, thus forbidding $\kap$ and $s$ from taking larger values.

The $H_1$ can play the role of $\hobs$ over the
complementary, and hence comparatively much larger, NMSSM parameter
space, where $\tb$ is not small simultaneously with large $\lam$. 
However, in the NNUHM case, a nested sampling of the NNUHM 
parameter space (as was done in our scans) tends to direct
the scans towards regions yielding $H_1$ solutions which are very
(MS)SM-like, owing to the collider constraints imposed, 
given that sufficiently large $\mzero$ and/or $\azero$ are allowed. 
In such regions, it is much harder to
obtain a $\neut{1}$ with $Z_s \geq 0.05$, since $\lam$ is typically
small and also since the relic density constraint is more 
readily satisfied by a gaugino or a higgsino or their admixture. 
In our scans of the NNUHM for the $\hobs=H_1$ scenario, we found 
only a handful of points with a non-vanishing singlino component.
We therefore carried out numerical scans of the more general NMSSM-14 
for this scenario only, in order to obtain a larger number of points 
for which $\neut{1}$ had a varying but
sizeable singlino component. The scanned ranges of the NMSSM-14 
parameters are given in table~\ref{tab:params}(b). 

\begin{table}[tbp]
\begin{center}
\begin{tabular}{cc}
\begin{tabular}{|c|c|}
\hline
NNUHM parameter & Scanned range  \\
\hline
$\mzero$\,(GeV) 	& 500 -- 2000 \\
$\mhalf$\,(GeV)  & 300 -- 1000	\\
$\azero$\, (GeV)  & $-5000$ -- 0\\
$\tb$ 		& 1 -- 25 \\
$\lam$ 		& 0.15 -- 0.7 \\
$\kap$ 		& 0.001 -- 0.4 \\
$\mue$\,(GeV) 	& 100 -- 200 \\
$\alam$\,(GeV)  	& $-1000$ -- 10000 \\
$\akap$\,(GeV)  	& $-500$ -- 2500\\
\hline
\end{tabular}
&
\hspace*{1cm}\begin{tabular}{|c|c|}
\hline
 NMSSM-14 parameter & Scanned range  \\
\hline
 $M_{\widetilde{Q}_3}$\,(GeV) & 1000 -- 10000\\
$M_{\widetilde{U}_3}$\,(GeV) & 1000 -- 10000\\
$M_{\widetilde{D}_3}$\,(GeV) & 1000 -- 10000\\
$M_{\widetilde{Q}}$\,(GeV) & 2000 -- 10000\\
$M_{\widetilde{L}}$\,(GeV) & 500 -- 10000\\
 $M_1$\,(GeV) & 500 -- 10000\\
$M_2$\,(GeV)  & 500 -- 10000  \\
$A_0$\,(GeV)  &  $-25000$ -- 0 \\
$\mue$ \,(GeV) & 100 -- 2000 \\
$\tb$ & 1 -- 70  \\
$\lam$  & 0.001 -- 0.7   \\
$\kap$  & 0.001 -- 0.7 \\
$\alam$\,(GeV) & 0 -- 25000 \\
$\akap$\,(GeV) & $-25000$ -- 0  \\
\hline
\end{tabular} \\
(a) & \hspace*{1cm} (b) \\
\end{tabular}
\end{center}
\caption{Scanned ranges of the input parameters for (a) the NNUHM
with $\hobs=H_2$ and (b) the NMSSM-14 with $\hobs=H_1$.}
\label{tab:params}
\end{table}

\begin{figure}[t]
\centering
\begin{minipage}{0.49\linewidth}
\includegraphics[width=\linewidth]{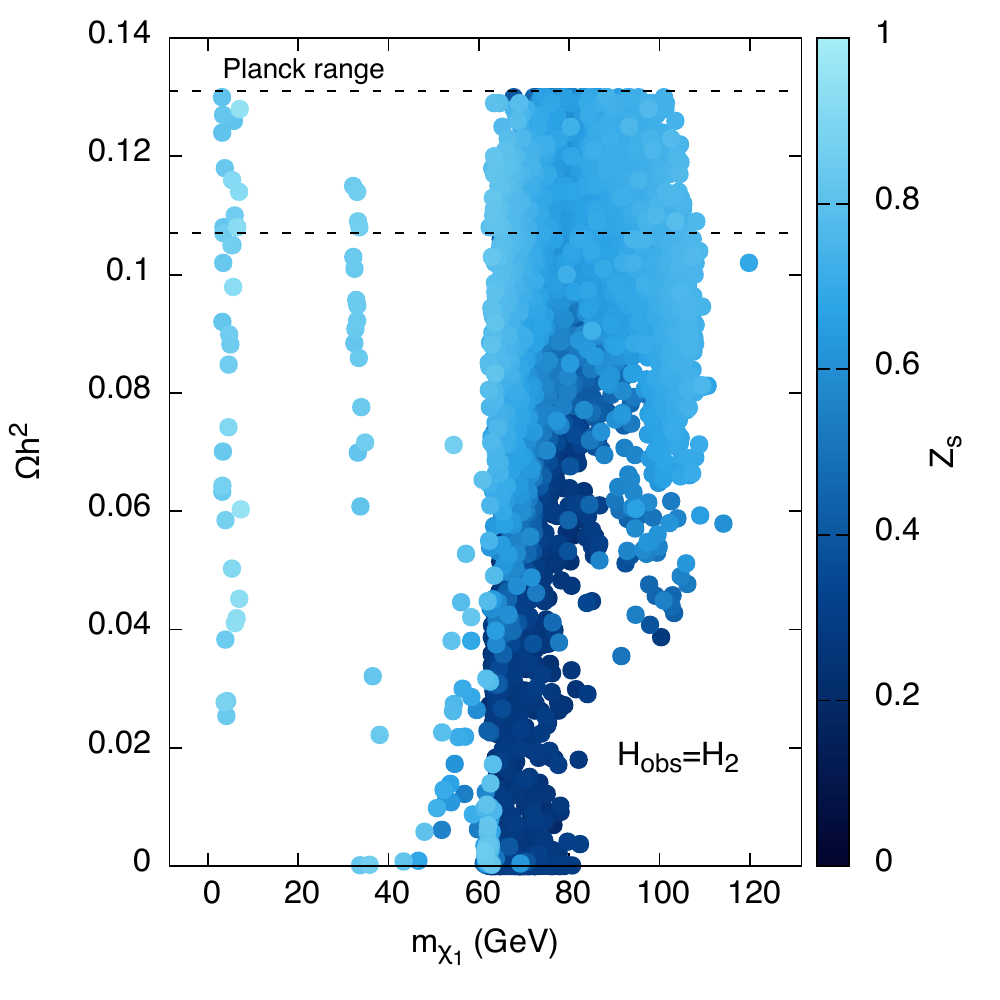}
\end{minipage}
\begin{minipage}{0.49\linewidth}
\includegraphics[width=\linewidth]{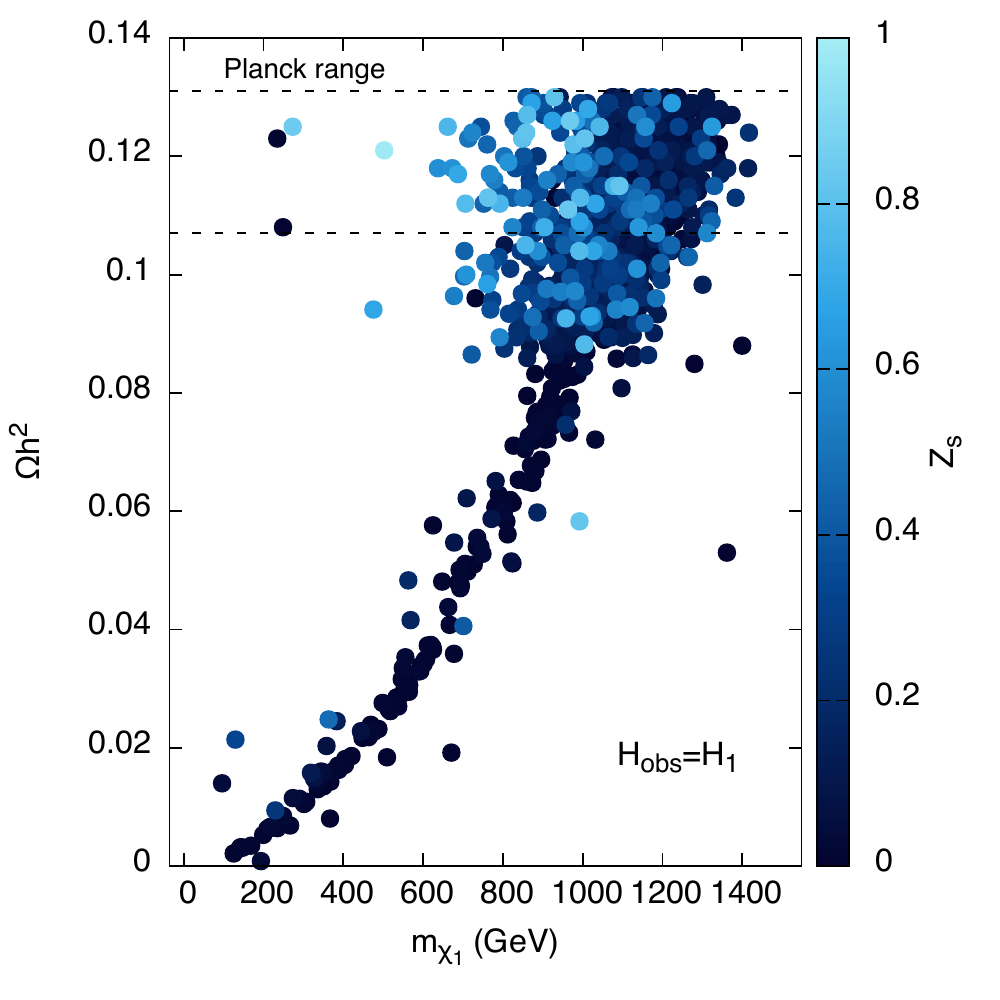}
\end{minipage}
\caption{
Relic density of the $\neut{1}$ as a function of its mass
for (a) the $\hobs=H_2$ case and (b) the $\hobs=H_1$ case.
The colour code corresponds to the singlino fraction.
}
\label{fig:compo}
\end{figure}

In figure~\ref{fig:compo}(a) we show $\Omega_{\neut 1} h^2$ as a function
of the $\neut{1}$ mass, $m_{\neut{1}}$, for the good points
obtained from our scan for the NNUHM with $\hobs = H_2$. 
The colour code in the figure shows the distribution of $Z_s$. 
The maximum $\Omega_{\neut 1} h^2$ seen for the points is 0.131, 
which allows for, as noted earlier, up to $+10\%$ theoretical error
given the Planck measurement of 0.119. A line corresponding to
the $-10\%$ error, at $\Omega_{\neut 1} h^2  = 0.107$, is also shown.
The points above this line are considered to be in agreement
with the Planck measurement and are henceforth we refer to as
 the `Planck-consistent' ones. We note an appreciable density of such points with 
$60\gev \lesssim m_{\neut{1}} \lesssim 100\gev$, with $Z_s$ for them 
increasing with decreasing
$m_{\neut{1}}$. These points are particularly interesting since, as
noted from figure~1 in\ \cite{Bergeron:2013lya}, $\neut{1}$ solutions
lying within this mass
range and also giving the correct $\Omega_{\neut 1} h^2$ are almost
unavailable in the MSSM. There are another two distinct sets of 
Planck-consistent points, one with $m_{\neut{1}} \simeq 32\gev$ and the other, 
also an NMSSM-specific one, with $m_{\neut{1}} < 10\gev$. 
$Z_s$ for these points is always very 
large. $m_{\neut{1}}$ in this scenario does not
exceed $\sim 120\gev$, for reasons discussed above. 

Figure~\ref{fig:compo}(b) similarly shows the distribution of
$m_{\neut{1}}$ for the good points from our NMSSM-14 scan for the
$\hobs = H_1$ scenario. Here also a very
interesting $m_{\neut{1}}$ range can be seen for the
Planck-consistent points. Neutralinos giving a good $\Omega_{\neut 1}
h^2$ and having a mass in the $\sim 500 -
1000\gev$ range are also nearly impossible to realise in the MSSM, 
again according to figure~1 in\ \cite{Bergeron:2013lya}. Similarly to the
$\hobs = H_2$ scenario above, for the Planck-consistent points the
singlino fraction increases as $m_{\neut{1}}$ drops, even though the
typical value of the latter is much larger. 
The large density of points with $m_{\neut{1}} \gtrsim 1\tev$ and a fairly small $Z_s$
generally is owing to the fact that a highly higgsino-dominated
neutralino can easily satisfy the $\Omega_{\neut 1} h^2$ constraint for such
masses. In fact none of the points with $m_{\neut{1}}\gtrsim 1.4\tev$
had $Z_s$ larger than 0.05, which is also evident from the figure. 

Overall, one notices in the two figures the trend 
that, for a given $m_{\neut{1}}$, a larger singlino fraction results in
larger $\Omega_{\neut 1} h^2$. This is expected, since a $\neut{1}$
with a small $Z_s$ (implying a large higgsino fraction)
has a higher interaction strength and hence a
higher annihilation rate compared to one with the same mass but 
a comparatively larger singlino fraction. There are thus some discrete
$\neut{1}$ mass ranges for which the $\Omega_{\neut 1} h^2$ obtained is consistent
with the Planck observation. In between these regions no Planck-consistent
neutralino solutions were obtained due to the fact that for the
corresponding masses the non-singlino fraction becomes too small to 
obtain sufficient DM (co-)annihilation. 

In the following, we shall not distinguish between 
points belonging to the NNUHM or to the NMSSM-14, since we are 
only concerned with the composition and the mass of a given $\neut{1}$. 
Whether a $\neut{1}$ corresponds to 
the $\hobs = H_1$ scenario or to the $\hobs =H_2$ scenario can be easily
inferred from its mass.

%-------------------------------------------------------------------------------------------------
\section{Dark matter in the Sun \label{sec:dmsun}}
%-------------------------------------------------------------------------------------------------

The description in this section follows the notation of\ \cite{Belanger:2013oya}. The number of neutralinos in the Sun as a function of time, $N_\chi(t)$, is given by an evolution equation\ \cite{Gould:1987ir}
\begin{align}
\frac{d N_\chi(t)}{dt} = C_\chi - A_{\chi\chi} N_\chi(t)^2 - E_\chi N_\chi(t)\,,
\label{eq:evolution}
\end{align}
where $C_\chi$ is the neutralino capture rate (assumed to be
constant),  $E_\chi N_\chi(t) $ is the evaporation rate, and
$A_{\chi\chi} N_\chi(t)^2$ is the annihilation rate of a pair of
neutralinos. Here $A_{\chi\chi}$ is defined as
\begin{align}
A_{\chi\chi} = \frac{\langle \sigma v \rangle}{V_\text{eff}^{DM}}\,,
\end{align}
where $\langle \sigma v \rangle$ is the averaged annihilation 
cross section times velocity, and 
$V_\text{eff}^{DM}$ is the effective volume of DM in the Sun\ \cite{Belanger:2013oya}.
Evaporation can be considered negligible for DM candidates above a few 
GeV\ \cite{Gould:1987ju}, in which case 
$N_\chi(t) \propto \tanh(\sqrt{A_{\chi\chi} C_\chi} t)$. Thus,
provided that the capture and annihilation rates are large 
enough, an equilibrium will be reached after a typical time scale of a few times $(A_{\chi\chi} C_\chi)^{-1/2}$, after 
which capture and annihilation balance each other out and the number of
the DM particles is roughly constant, 
$N_\chi\sim \sqrt{C_\chi/A_{\chi\chi}}$. The annihilation rate is then given by the capture rate alone,
$\Gamma_{\chi\chi}=A_{\chi\chi}N_\chi^2/2=C_\chi/2$. 
For the Sun, most models lead to equilibrium, but \MO{} does not 
assume this is the case and solves the evolution 
eq.~(\ref{eq:evolution}) numerically.
The capture rate depends on the cross section for WIMP--nucleus
elastic scattering and on the local density and 
velocity distribution in the Galaxy\ \cite{Gould:1987ir}. The
default value of the local DM density, $\rho_0$, in \MO{} is
0.3~GeV/cm$^3$ and a Maxwellian velocity distribution is assumed.  

The annihilation of a pair of WIMPs can lead to many different final
states which decay further and eventually yield neutrinos. 
The energy spectrum of the neutrinos produced in the centre of the Sun 
is computed by \MO{} using the results of\ \cite{Cirelli:2005gh},
for each point in the NMSSM parameter space considered. The 
resulting neutrino flux at the Earth is then obtained by \MO{} by
computing the appropriate BRs using the NMSSM model 
(which in turn calls \NMT). The number of events expected in 
IceCube is then calculated by convoluting this 
flux with the detector effective area and angular resolution, 
as explained in the following sections. 

%-------------------------------------------------------------------------------------------------
\section{The IceCube and PINGU detectors \label{sec:icecube}}
%-------------------------------------------------------------------------------------------------

The IceCube neutrino detector\ \cite{Achterberg:2006md} at the South Pole consists of one km$^3$ of very clear ice instrumented 
with 5600 optical modules arranged in 86 vertical strings. The optical modules register the Cherenkov radiation emitted by 
charged leptons produced in high-energy neutrino interactions in the ice.
For the purpose of this paper, we are particularly interested in muons
from the interactions of muon neutrinos, because such 
muons are observed as tracks giving a good angular resolution. Although the IceCube geometry has been optimised to detect 
ultra-high energy ($> \tev$) neutrinos from potential cosmic sources with sub-degree angular resolution, its angular 
response of a few degrees at $\mathcal{O}$(100\,GeV) is still adequate to perform directional searches for DM. 
Unfortunately the energy of muon-neutrino induced events can not be precisely 
measured since the long muon track might not be contained within the detector volume. Electron neutrinos and tau neutrinos, 
on the other hand, leave localised energy depositions in the detector that, while 
offering much worse angular resolution, allow for a better energy measurement since they are contained in the detector. 
However, since we are interested in neutrinos originating in the centre of the Sun, it is important to be able to identify 
the direction, and we are therefore going to consider only muon neutrinos in the following. In the centre of IceCube there 
is a more densely instrumented sub-array called DeepCore, which is more sensitive to lower energy neutrinos, having an 
energy threshold around 10 GeV\ \cite{Collaboration:2011ym}. As we will see below, it is very important to have 
sensitivity to low-mass DM. In addition to the IceCube and DeepCore detectors, we consider a proposed extension called 
PINGU, a low-energy in-fill of about 40 additional strings with closer optical module spacing, which will 
have an energy threshold for neutrinos of a few GeV\ \cite{Aartsen:2014oha}.

The efficiency of a neutrino telescope to a given signal is parametrised by the effective area for muon neutrinos, $A_{\nu_\mu}^\text{eff}$, which is 
a function of the neutrino energy. The effective area is the equivalent area where the detector can detect a neutrino with 
100\% efficiency, and includes the neutrino-nucleon interaction probability, the energy loss of the produced muon from 
the interaction point to the detector and the detector trigger and analysis efficiency.  For the IceCube/DeepCore analysis we use the detector 
effective areas presented in\ \cite{DanningerThesis2011}, which were obtained from detailed simulations of the detector response within the 
framework of a search of IceCube data for DM in the Sun\ \cite{Aartsen:2012kia}. Below 40~GeV we use the 
PINGU effective area presented in~\cite{Aartsen:ICRC2015a}. We then have three practically independent
detectors, together covering a large neutrino energy range, with
little overlap between their individual coverages, as shown in figure~\ref{fig:spectra-examples} below.

\section{Signal and background estimations \label{sec:sb}} 
\subsection{Expected signal from the Sun \label{sec:signal}} 

We want to calculate the number of events that IceCube, DeepCore and
PINGU can expect for the different NMSSM model points we analyse, and compare it 
with the number of expected background events from the direction of the Sun.
The neutrino flux $\Phi$, obtained using \MO{} from the NMSSM model,
is parametrised as
\begin{equation}
\Phi = \frac{dN_{\nu_\mu}}{dE_{\nu_\mu} \, dA \, dt \, d\Omega}\,,
\end{equation}
with neutrino energy $E_{\nu_\mu}$, area on the Earth $A$, time $t$
and solid angle $\Omega$ of the sky.
$\Omega$ is defined in terms of the polar angle $\theta$ and the azimuthal angle
$\varphi$ as $d\Omega = d\varphi \, d\theta \, \sin\theta$. \MO{} returns the
neutrino and anti--neutrino fluxes separately. Even if we take ``neutrino flux'' to
denote both neutrino and anti--neutrino flux in what follows, we treat neutrinos and
anti--neutrinos separately in our calculations and use the corresponding cross sections when needed.
Even though the signal from DM annihilation is independent of $t$,
the background depends on the position of the Sun and, therefore, on $t$.
For simplicity, we choose the coordinate system with $\theta=0^\circ$ in 
the direction of the Sun. For a given detector, the total number of
events observed from an incident neutrino flux is given by
\begin{equation}
N_{\nu_\mu} =
	\int dt
	\int_{0}^\infty dE_{\nu_\mu}
	\int_0^{2\pi} d\varphi
	\int_0^{\theta_\mathrm{cut}} d\theta \sin\theta
	A_{\nu_\mu}^\text{eff}(E_{\nu_\mu}) \, \Phi(E_{\nu_\mu},\theta,\varphi,t)
	\,,
\label{eq:nevents}
\end{equation}
where $A_{\nu_\mu}^\text{eff}$ is the effective area. 
The integral over $t$ is done over the live time of the detector, as
will be specified later. In our case of a search for DM annihilation in the Sun,
the angular integral is performed over a cone that points to the Sun's position
with a cut on the maximum polar angle $\theta_\mathrm{cut}$
to take into account the angular resolution of the detector.
$\Phi$ at the detector location (Earth) will be modified by
neutrino energy losses as well as oscillations
on their way from the Sun. \MO{} takes care of these effects,
providing the expected neutrino flux at the Earth for 
each signal model point considered.
In a similar way, events from anti-neutrinos are calculated.

%In the case of PINGU, for which the detector effective volume is
%available instead of the area, the effective area is obtained via
%\begin{equation}
%A_{\nu_\mu}^\mathrm{eff} = \frac{V_{\nu_\mu}^\text{eff}(E_{\nu_\mu}) \, \sigma_{\nu N}(E_{\nu_\mu}) \, \rho_\text{ice} \, A_\text{ice}\, N_A}{M_\text{ice}} \,,
%\end{equation}
%\label{eq:nevpingu}
%where $\rho_\text{ice}$, $A_\text{ice}$ and $M_\text{ice}$ are the density, atomic number and molar mass of the ice, respectively.
%We take the density of ice at the South Pole to be $0.92$~g/cm$^3$, the atomic number $18$ and the molar mass $18.015$~g/mol.
%$N_A$ is the Avogadro's number, and $\sigma_{\nu N}$ is the
%cross section for charged current deep inelastic scattering of the
%incident neutrinos with nucleons in the ice.  We use the cross sections 
%from\ \cite{Gandhi:1995tf,Kumar:2011hi}.

%------------------------------------------------------------------------------------------------
\subsection{Atmospheric background \label{sec:background}}
%-------------------------------------------------------------------------------------------------

 Neutrinos created in the atmosphere by cosmic ray interactions 
(see e.g.~\cite{Volkova:1983yf,Gaisser:2002jj,Barr:2004br,Honda:2006qj}) 
constitute an irreducible background to the signal, all year round. On top
of this background, atmospheric muons produced in the same 
cosmic ray interactions provide an additional background that can be
dealt with differently depending on the time of the year. 
During the antarctic summer season, when the Sun is above the horizon, 
neutrinos from the Sun are \textit{downgoing}. 
Any search for a solar signal is then plagued by the overwhelming
atmospheric muon background, which is about seven orders of magnitude larger 
than the atmospheric neutrino background, depending on the 
declination~\cite{Gaisser:2002jj,Honda:2006qj}. It is only through vetoing techniques 
(considering events starting inside the detector volume,
which~\textit{must} be 
caused by a neutrino) that IceCube has managed to perform searches 
for DM in the Sun in the austral summer also. The price to pay is a
significantly reduced detector volume, since part of the outer strings of 
the detector are used for tagging incoming atmospheric muons. Such
an analysis requires access to detailed detector event information, which is beyond the
scope of this work. We will thus concentrate on a ``winter''-type
analysis in the rest of the paper by using the corresponding effective areas 
from\ \cite{DanningerThesis2011}. 

 During the antarctic winter season, the Sun is below the horizon at
 the South Pole and neutrinos coming from it are \textit{upgoing}, 
meaning that they pass through the Earth before arriving at IceCube. 
This provides a filter against the atmospheric muon 
background. The only physical background then is the atmospheric neutrino flux. There is, however,
an additional background component consistent of downgoing atmospheric
muons misreconstructed as upgoing. This is a background that can be minimised 
by clever data analyses, but a residual contamination of a few percent
can always be present in the final analysis. We will assume in the 
rest of the paper that such misreconstructed atmospheric muon
contamination can be reduced to an insignificant level in a real event-by-event 
analysis, and we will not consider such a background further in our
calculations. Our results are then slightly optimistic, but we note that a few 
percent increase, $\epsilon$, in the background would worsen the
significances presented in section~\ref{sec:stat} just at the level of 1/$\sqrt{1+\epsilon}$.

For our estimations of the PINGU sensitivity, on the other hand, we
will not make any distinction between winter and summer periods. This
is because we assume 
that using the surrounding IceCube detector as a veto, downgoing 
atmospheric muons can be 
effectively rejected in the summer period and the detector can be used 
all year round in the same way.

 We calculate the expected background from state-of-the-art atmospheric neutrino flux calculations. 
At the energies relevant for our analysis, this background is
dominated by the conventional neutrino flux from decays of pions and kaons. 
The prompt flux from charmed hadrons is only relevant for much higher
energies. The conventional flux depends on the neutrino energy and the 
zenith angle. The flux prediction by Honda et al.\ \cite{Honda:2006qj}
has been modified by Gaisser\ \cite{Gaisser:2012zz} to account for the 
cosmic ray knee and agrees well with the measured flux. It is used by
the IceCube collaboration as a theoretical standard prediction for the conventional 
flux. We use this flux for our background estimates through the
implementation in the code NeutrinoFlux~\cite{atmospheric}. The calculation proceeds 
along the same lines as for the signal estimation, that is, through
the convolution of the atmospheric neutrino flux with the effective area/volume of 
the detector as in~\refeq{eq:nevents}. 

\begin{figure}
\centering
\begin{minipage}{0.99\linewidth}
\includegraphics[width=\linewidth]{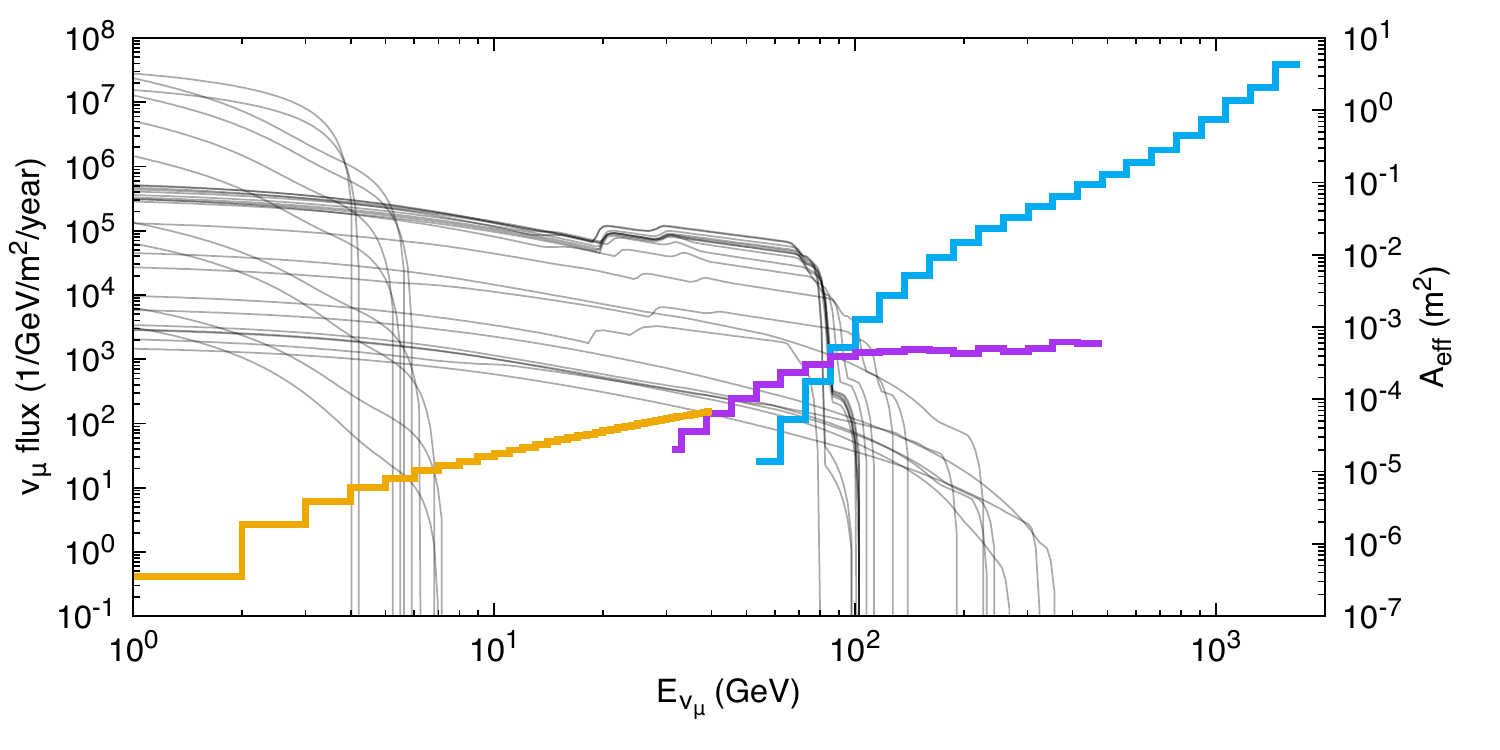}
\end{minipage}
\caption{
Examples of the $\nu_\mu$ spectra (left axis) from some selected model points which
result in a large statistical significance.
The non-smooth features in some of the spectra originate from DM annihilation into $WW$ or $ZZ$.
Also shown is the energy dependence of the
PINGU (orange), IceCube (blue) and DeepCore (violet) effective areas (right axis).
}
\label{fig:spectra-examples}
\end{figure}

%-------------------------------------------------------------------------------------------------
\section{Results and discussion \label{sec:stat}}
%-------------------------------------------------------------------------------------------------

\begin{figure}
\centering
\begin{minipage}{0.49\linewidth}
%\pdfoptionpdfminorversion=5
\includegraphics[width=\linewidth]{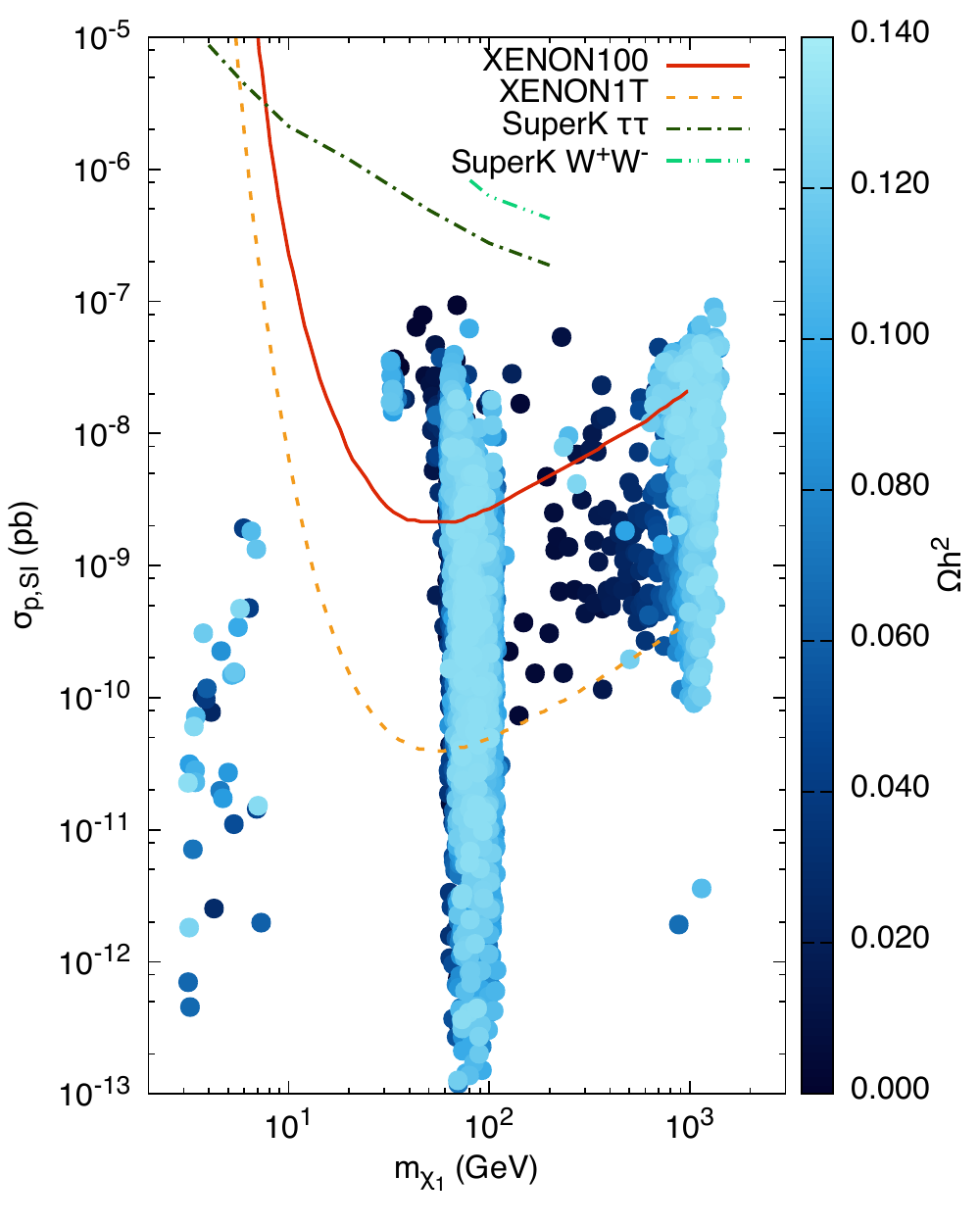}
\end{minipage}
\begin{minipage}{0.49\linewidth}
\includegraphics[width=\linewidth]{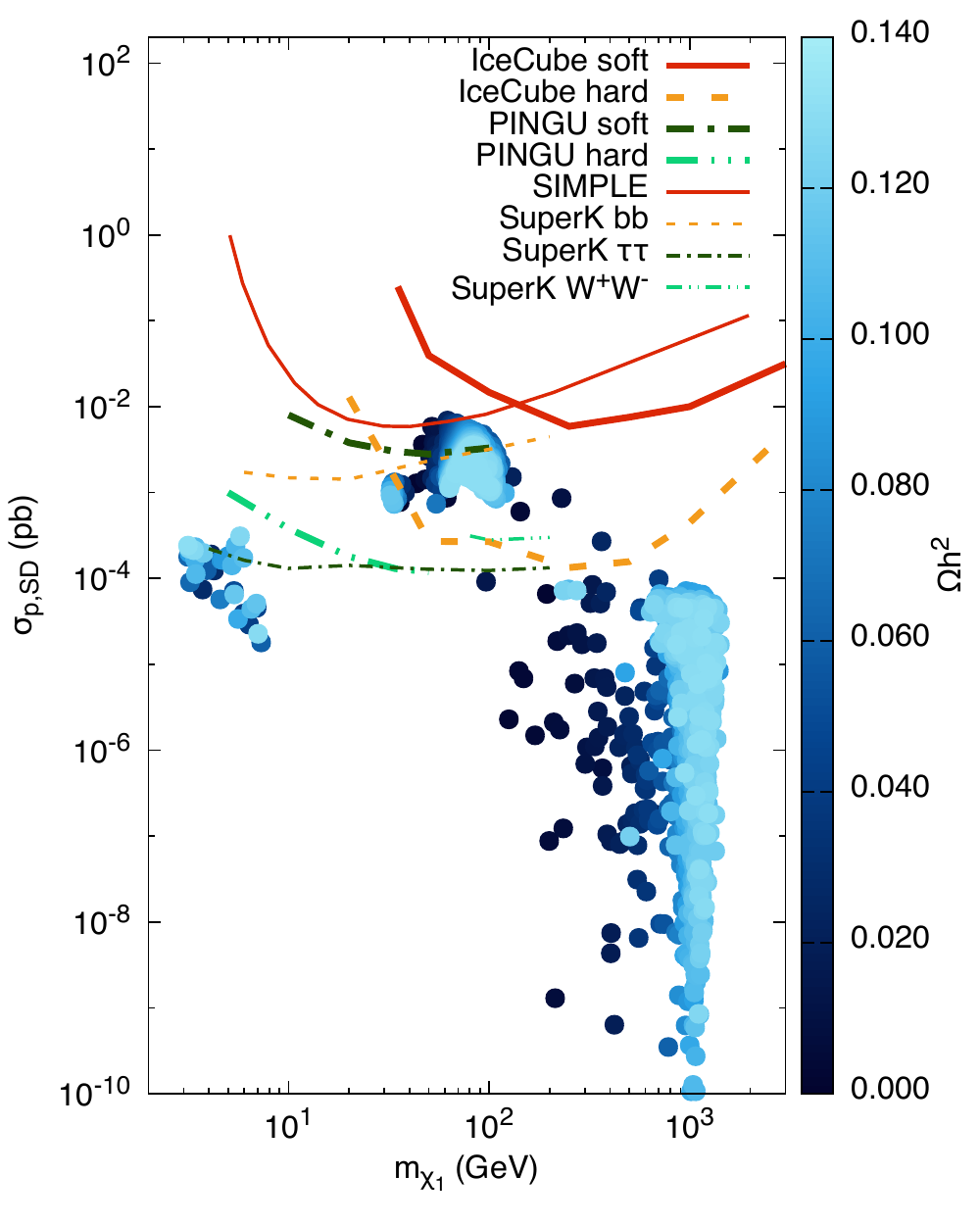}
\end{minipage}
\\
\begin{minipage}{0.49\linewidth}
%\pdfoptionpdfminorversion=5
\includegraphics[width=\linewidth]{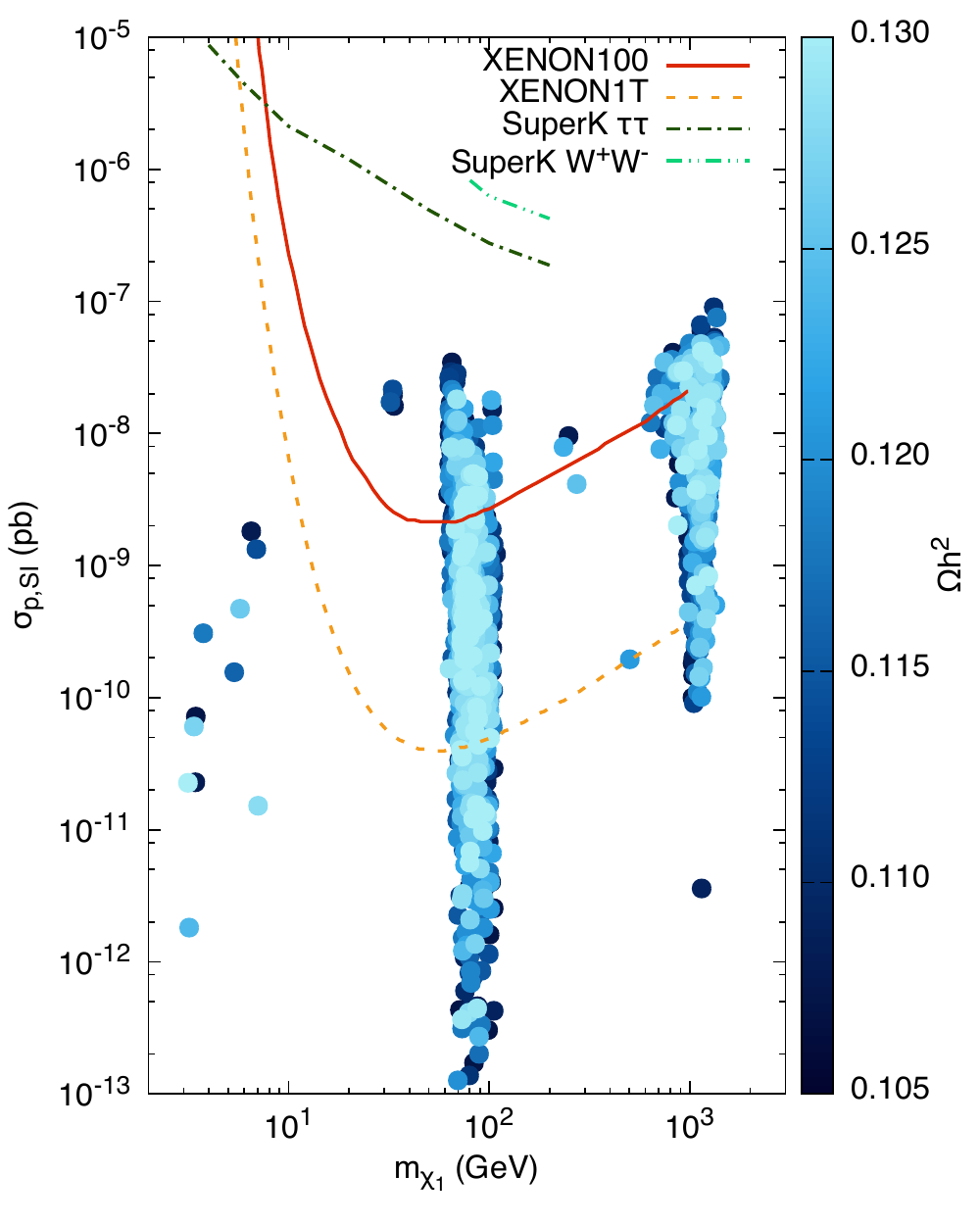}
\end{minipage}
\begin{minipage}{0.49\linewidth}
\includegraphics[width=\linewidth]{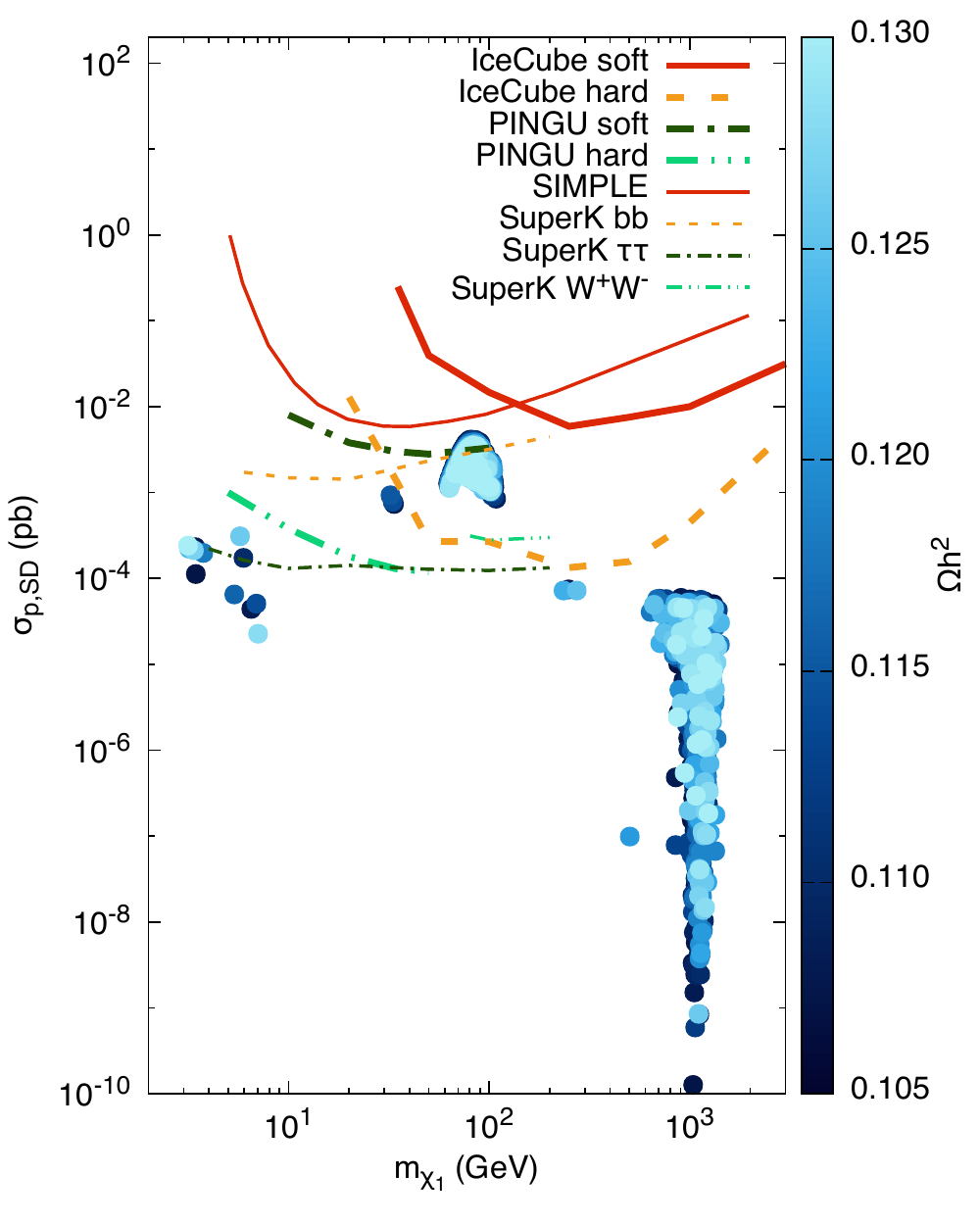}
\end{minipage}
\caption{
Scatter plots of the model points showing $\sigma_{p, SI}$ (left)
and $\sigma_{p, SD}$ (right) in terms of the $\neut{1}$
mass, with the colour code corresponding to $\Omega_{\bar{\chi}^0_1}h^2$.
The top panels contain all the good points from the scans, while
the lower panels show only the model points with the predicted 
$\Omega_{\bar{\chi}^0_1}h^2$ lying within
$\pm 10\%$ of the Planck measurement.}
\label{fig:m_neut1-sigsip-sigsdp-oh2--relic-OK-DDOK-sip}
\end{figure}

The predicted $\nu_\mu$ spectra from DM annihilations in the Sun are obtained from the NMSSM model points via \MO,
and the expected signal is calculated using the effective areas of the detectors as indicated in eq.~(\ref{eq:nevents}). 
We consider the three independent detectors IceCube, DeepCore and
PINGU, as noted above and illustrated in figure~\ref{fig:spectra-examples}, 
each covering a different range in neutrino energy. Figure~\ref{fig:spectra-examples} also contains examples of the
$\nu_\mu$ spectra resulting from DM annihilation in the Sun. A few spectra with the largest signal were selected to 
illustrate the general features of the distributions. The irregular structure of some of the spectra around \unit[30]{GeV}
is due to the annihilation channel $\neut{1} \, \neut{1}  \to W^+W^-$,
which can result in such characteristic features if the DM is heavy enough so that this
channel is possible, but light enough for the spectrum to not get
smoothed out\ \cite{Cirelli:2005gh}.

For a predicted number of signal events $N_s$ from a model and a given number of background events $N_b$, we
estimate the significance $S$ by the formula\ \cite{Cowan:2010js}
\be
S = \sqrt{2\left( (N_s+N_b) \ln\left(1+\frac{N_s}{N_b}\right)-N_s \right)} \approx
\frac{N_s}{\sqrt{N_b}}\,,
\label{eq:significance}
\ee
where the last approximation is valid for $N_s\ll N_b$.
The complete expression above yields a lower $S$ compared to the
approximate one in the case of large $N_s$ and the difference between
the two amounts in our case to a maximum correction of 20\%.

We integrate the signal and background over the actual live times of the detectors during one calendar year.
As pointed out earlier, PINGU is assumed to take data during the whole
year, since we assume
that IceCube can be used as an efficient veto.
IceCube and DeepCore are assumed to take data only during 152 days of
the winter time as specified in table~7.1 of~\cite{DanningerThesis2011}.
This lets us compare the performance of the different detectors in a more fair way.

We cut on the angle $\theta$ between the reconstructed track and the direction of the Sun
as outlined in eq.\ (\ref{eq:nevents})
to take into account the reconstruction error for the neutrinos from the point--like source.
For the signal, we assume that $\theta$ is normal--distributed around $0^\circ$
with a median given by figure~7.17 in \cite{DanningerThesis2011}.
$S$ is then maximised over $\theta_\mathrm{cut}$, with the typical optimum values being $3^{\circ}$ for IceCube and
$8^{\circ}$ for DeepCore, while we use $20^{\circ}$ for PINGU.
The exact optimum value depends also on the specific NMSSM model
point, since the angular resolution depends on the neutrino energy.

In the context of the current direct and indirect DM searches,
figure~\ref{fig:m_neut1-sigsip-sigsdp-oh2--relic-OK-DDOK-sip} shows
the spin-independent and 
spin-dependent DM-proton elastic scattering cross sections, denoted by
$\sigma_{p, SI}$ and $\sigma_{p, SD}$, respectively, as functions 
of $m_{\neut{1}}$, with the colour-code indicating the value of $\Omega_{\neut
  1} h^2$ predicted for each model point. 
The top panels show the complete set of good points obtained from all
our scans, while 
the lower panels contain only the Planck-consistent
(i.e., with $0.107 < \Omega_{\neut  1} h^2 < 0.131$) subset of these
points. The experimental 
limits shown in the left panels correspond to the current 90\% CL exclusion
limits from XENON100~\cite{Aprile:2012nq} as well as the projected
90\% CL limits for XENON1T~\cite{Aprile:2012zx}. In the right panels
we similarly show the 90\% CL exclusion limits from SIMPLE~\cite{Felizardo:2011uw}
and IceCube~\cite{Aartsen:2012kia},
as well as the projected 90\% CL exclusion limits for PINGU~\cite{Aartsen:2014oha}.
The current limits from IceCube are shown as a guideline, but they
should be interpreted with care. 
They were obtained under some simplifying assumptions - 100\%
annihilation rate into either $b\bar{b}$ (soft) or $W^+W^-$
(hard) - and therefore can not be
taken to exclude all model points lying above the corresponding
lines. A real NMSSM model point has a mixture of annihilation channels and these limits 
would be relevant for a parameter point only if it predicts a
neutralino annihilation rate almost exclusively into either $b\bar{b}$ or $W^+W^-$.

As can be seen from this figure, a fraction of the considered
parameter points with $m_{\neut{1}} \sim 100$\,GeV and $m_{\neut{1}} \sim 1$\,TeV
is already excluded by XENON100, and a large fraction of the remaining such
points can be tested with the upcoming XENON1T experiment. The XENON1T and
IceCube exclusions are, however, complementary, which we will come back to below. 
A note of caution here is the well-known fact that the direct and
indirect detection limits on SUSY models are subject to different
uncertainties from particle and nuclear physics and
astrophysics. In particular, the direct detection limits are affected
by large uncertainties from nuclear
physics~\cite{Bottino:1999ei,Ellis:2008hf,Crivellin:2013ipa}
(for recent analyses of the impact of such uncertainties on the
cMSSM and MSSM parameter spaces see,
e.g.,~\cite{deAustri:2013saa,Fowlie:2013oua,Crivellin:2015bva}). Indirect searches for
DM in the Sun are affected by several
astrophysical uncertainties, mainly on the value of the local DM density, the assumed velocity distribution 
and velocity dispersion, although they have a more moderate effect on the final limits~\cite{Choi:2013eda,Danninger:2014xza}. 
Therefore, while we consider the points that lie above the line corresponding to XENON100 in figure~\ref{fig:m_neut1-sigsip-sigsdp-oh2--relic-OK-DDOK-sip}
to be excluded, we shall anyhow include them in our discussion below, but separately from the allowed points, in order to 
illustrate the full complementarity of the direct and indirect searches.

\begin{figure}
\centering
\begin{minipage}{0.49\linewidth}
\includegraphics[width=\linewidth]{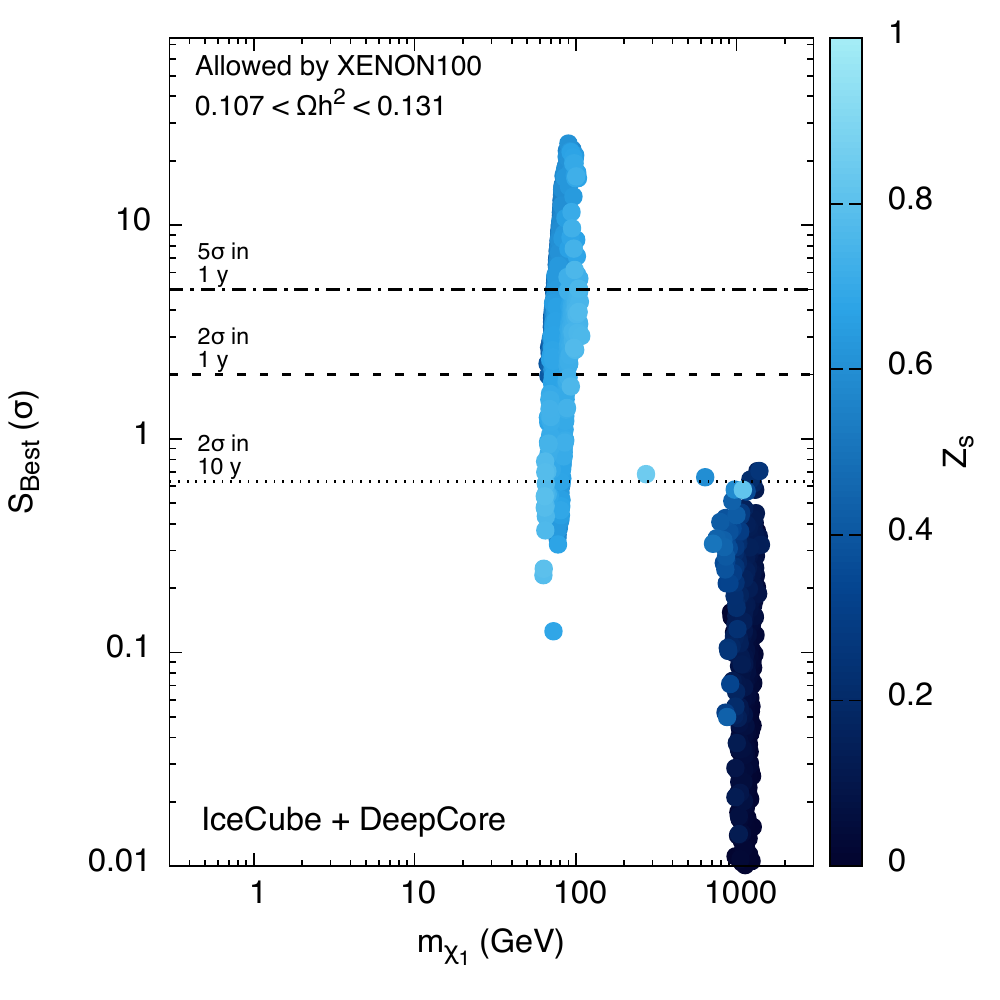}
\end{minipage}
\begin{minipage}{0.49\linewidth}
\includegraphics[width=\linewidth]{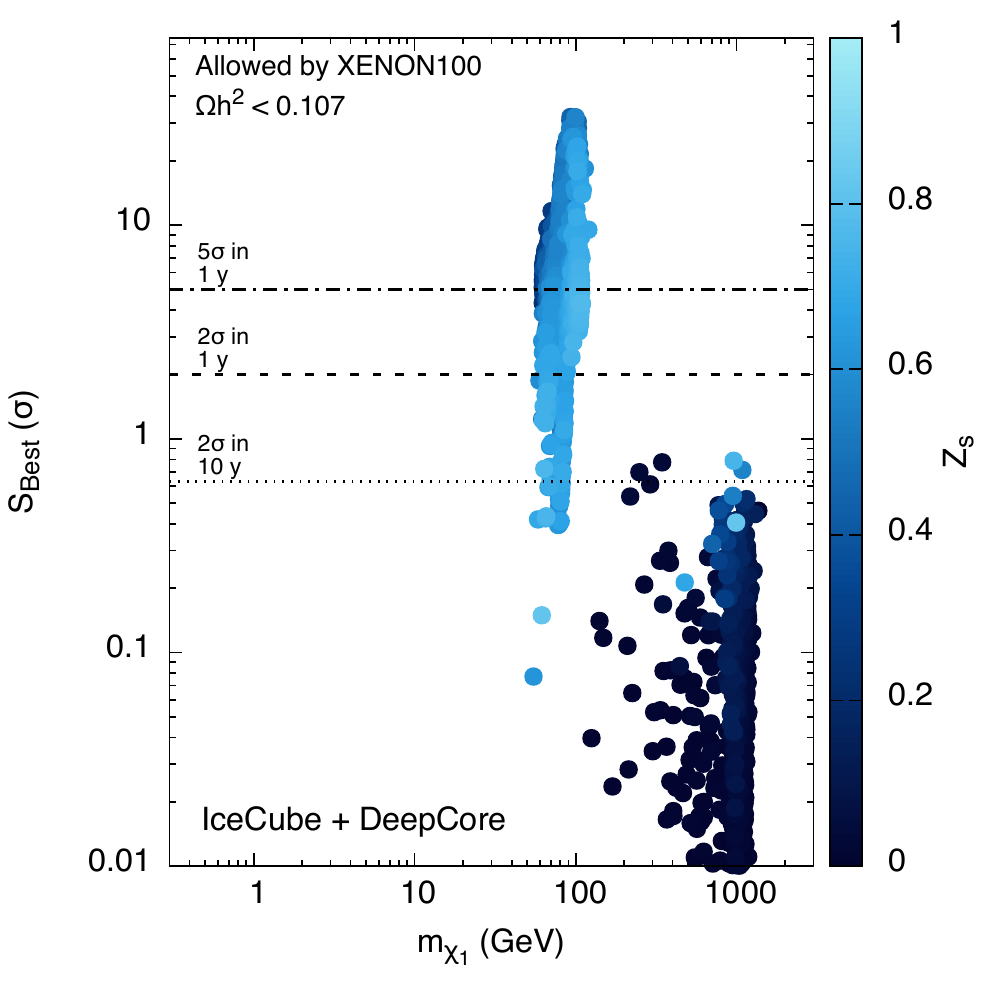}
\end{minipage}
\begin{minipage}{0.49\linewidth}
\includegraphics[width=\linewidth]{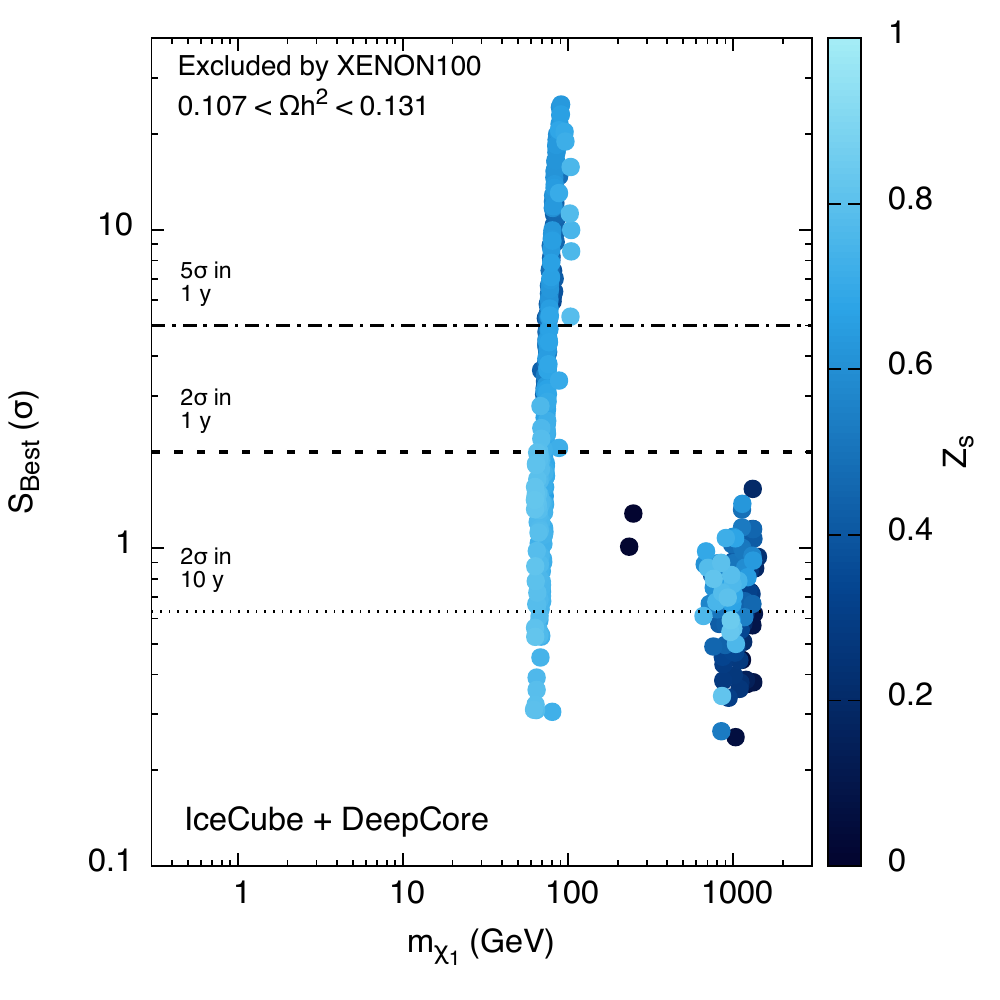}
\end{minipage}
\begin{minipage}{0.49\linewidth}
\includegraphics[width=\linewidth]{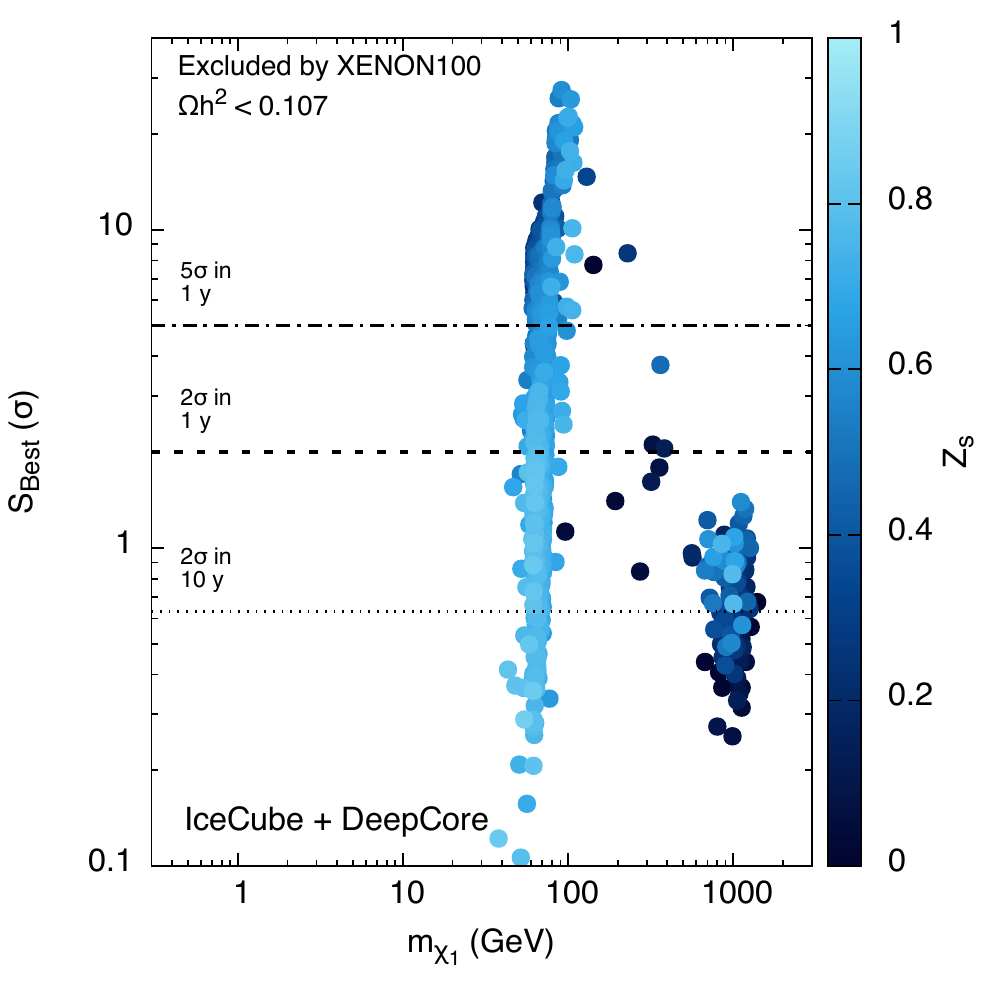}
\end{minipage}
\caption{
Overview of the maximum significance for each model point
obtainable from the IceCube or DeepCore detector.
The cut $\theta_\mathrm{cut}$ on the angle between the tracks and the
direction of the Sun is optimised for each model point.
The values for the significance represent 1 calendar year of data-taking.
IceCube and DeepCore are assumed to operate only during
152 days of winter, as described in the text,
but PINGU, shown in later figures, operates independent of season.
The colour code indicates the singlino fraction.
}
\label{fig:signi_figures_01}
\end{figure}

\begin{figure}
\centering
\begin{minipage}{0.32\linewidth}
\includegraphics[width=\linewidth]{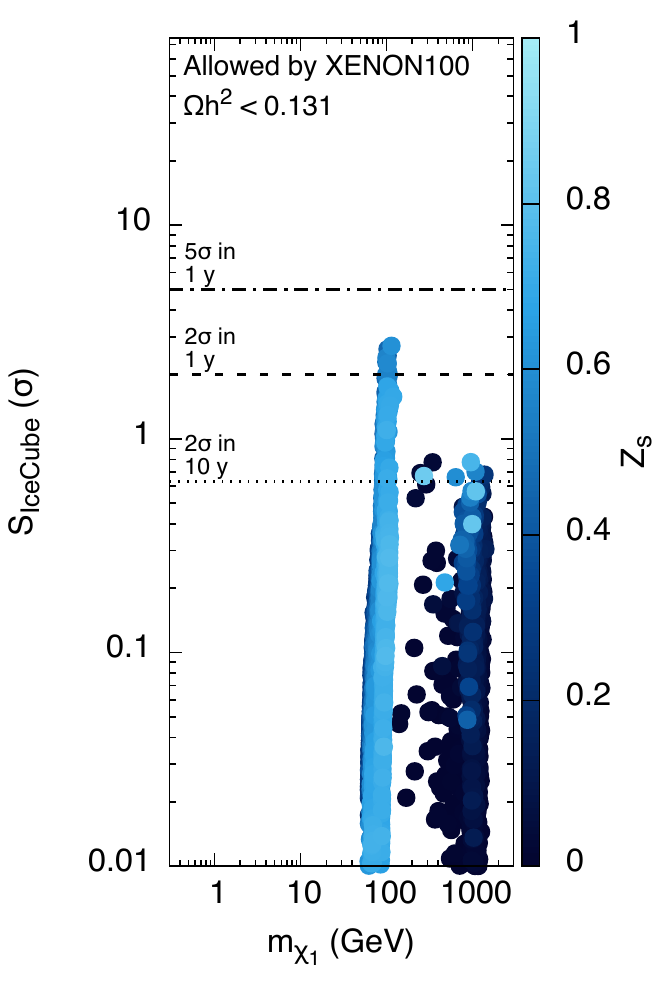}
\end{minipage}
\begin{minipage}{0.32\linewidth}
\includegraphics[width=\linewidth]{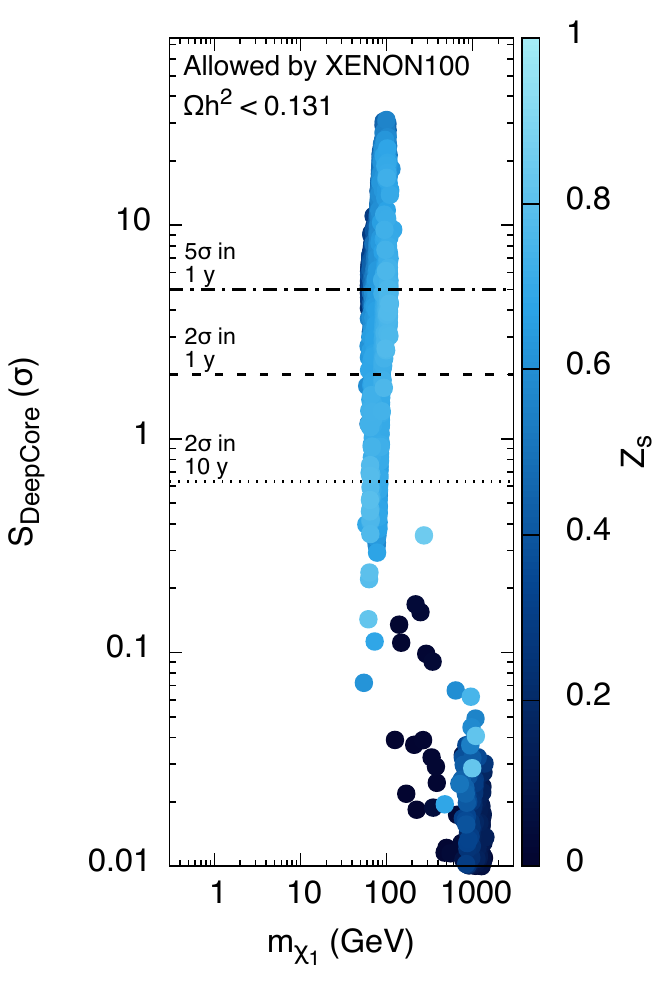}
\end{minipage}
\begin{minipage}{0.32\linewidth}
\includegraphics[width=\linewidth]{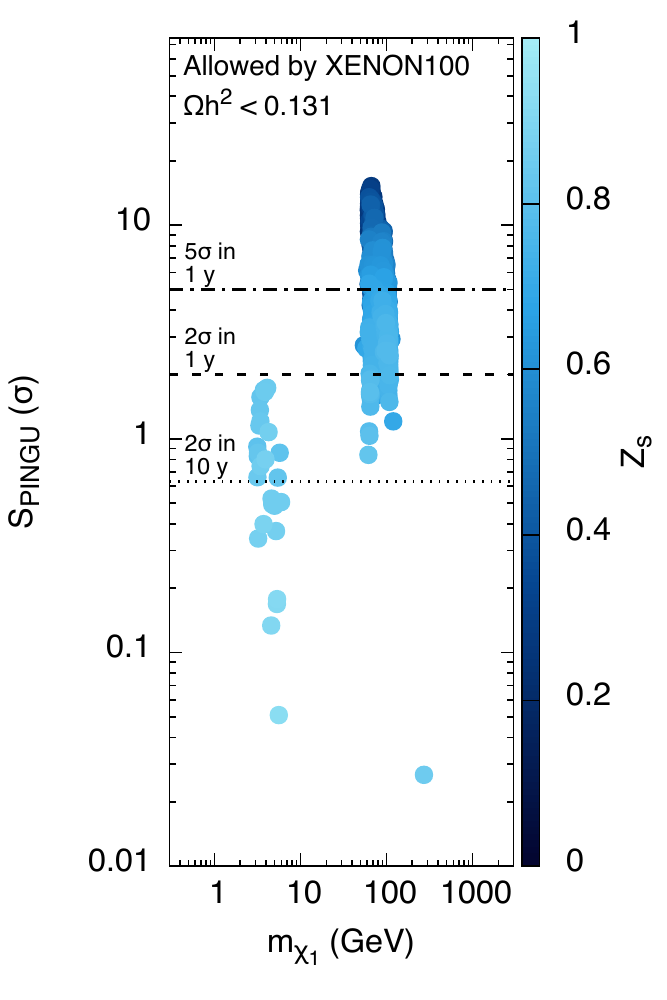}
\end{minipage}
\caption{
Expected significances at the IceCube (left), DeepCore (middle) and
PINGU (right) detectors.
We note that DeepCore already rules out a large fraction of the
parameter space points with $m_{\neut{1}} \sim 100\,\mathrm{GeV}$
with one calendar year of data-taking (cf.~figure~\ref{fig:signi_figures_01}),
and PINGU has the potential to reach lower-mass points than those accessible with DeepCore 
within a few years of runtime. All points shown pass the XENON100 limits.
}
\label{fig:signi_figures_02}
\end{figure}

The impact of IceCube, DeepCore and PINGU on the higgsino-singlino DM predictions in
the NMSSM is illustrated in figures~\ref{fig:signi_figures_01}  
and~\ref{fig:signi_figures_02}.
The figures show the significance $S$ defined in eq.~(\ref{eq:significance})
as a function of the $\neut{1}$ mass for the NMSSM model points after 1 calendar year.
Also drawn in the figures are the lines corresponding to discovery 
(5$\sigma$) as well as to exclusion (2$\sigma$) after 1 or 10
calendar years, the latter aiming to highlight the model points that
have not been tested by any of
the detectors yet but will be after 10 calendar years of data-taking. Note that the 
models above the 5$\sigma$ line in the plots would have probably been 
seen in the IceCube/DeepCore analysis 
presented in~\cite{Aartsen:2012kia} which, even when optimised for
 the MSSM $\neut{1}$, does not show any other specific 
model dependencies, and is general enough to be sensitive to the NMSSM
$\neut{1}$ as well. Such points are thus in principle already ruled
out. The colour code in both the figures shows $Z_s$, and
the model points have been sorted in such a way that a point with
larger $Z_s$ lies on top of another one with the same $m_{\neut{1}}$ and $S$.

Figure~\ref{fig:signi_figures_01} shows the best significance,
$S_\mathrm{Best}$, obtained for a given model point by either IceCube or DeepCore.
Results for PINGU will be shown later in a separate figure
because PINGU has not yet started to collect data.
The left panels contain only the Planck-consistent points while the right panels contain the
Planck-inconsistent ones, i.e, for which $\Omega_{\neut  1} h^2<0.107$.
The top panels show points that are not excluded by XENON100,
while the lower panels contain the complementary set, i.e., models 
disfavoured by XENON100.
One sees in the figures a high sensitivity to a large number of points with
$m_{\neut{1}} \sim \unit[100]{GeV}$,
which can partly be traced back to the DM annihilation
channel ${\neut{1}}{\neut{1}} \to W^+W^-$ yielding a harder $\nu_\mu$
spectrum for such a $\neut{1}$ mass~\cite{Cirelli:2005gh}. In fact, a large fraction
of the $m_{\neut{1}} \sim \unit[100]{GeV}$ points in each panel lie 
above the $2\sigma$ line shown. The top panels are of particular
interest in this regard, since they indicate that many of these
points, potentially ruled out by IceCube after just one calendar year
of data-taking, were not probed by the XENON100 experiment.  

Figure~\ref{fig:signi_figures_02}, on the other hand, shows the
performance of each detector individually, but for model points
that satisfy only the upper limit on $\Omega_{\neut{1}} h^2$ and are
allowed by XENON100. We see that 
it is DeepCore (middle) and PINGU (right)
which contribute more strongly to rejecting a large
number of the $m_{\neut{1}} \sim \unit[100]{GeV}$ points.
As can be seen in figures~\ref{fig:spectra-examples}
and~\ref{fig:m_neut1-sigsip-sigsdp-oh2--relic-OK-DDOK-sip}, the
${\neut{1}}{\neut{1}} \to W^+W^-$ annihilation channel is well covered by the
DeepCore detector for this $m_{\neut{1}}$ range.
Almost all the model points belonging in the 
$\unit[200]{GeV} \lesssim m_{\neut{1}} \lesssim \unit[1400]{GeV}$ strip
noted in the left panel (for IceCube) of figure~\ref{fig:compo}(b) lie
just below the line
corresponding to $2\sigma$ significance after 10 years and will thus 
not be accessible within this timescale. 
Note that the inclusion of PINGU in our analysis is mainly to estimate its
sensitivity to the model points with $m_{\neut{1}} \lesssim \unit[20]{GeV}$. 
The results show that PINGU is indeed sensitive to $\neut{1}$ in the
$\mathcal{O}(\unit[10][GeV])$ mass range, although it needs a few
years of accumulated livetime. Dedicated event-based 
likelihood analyses, on the lines of the one presented in~\cite{Scott:2012mq}, are 
known to improve the sensitivity of an experiment since they use more information 
in a more comprehensive way, and can enhance the prospects for reaching such points with 
higher significance in a shorter timescale. To evaluate the effect of such an approach 
is, however, beyond the scope of this paper. 

\begin{figure}
\begin{center}
\begin{minipage}{0.49\linewidth}
\includegraphics[width=\linewidth]{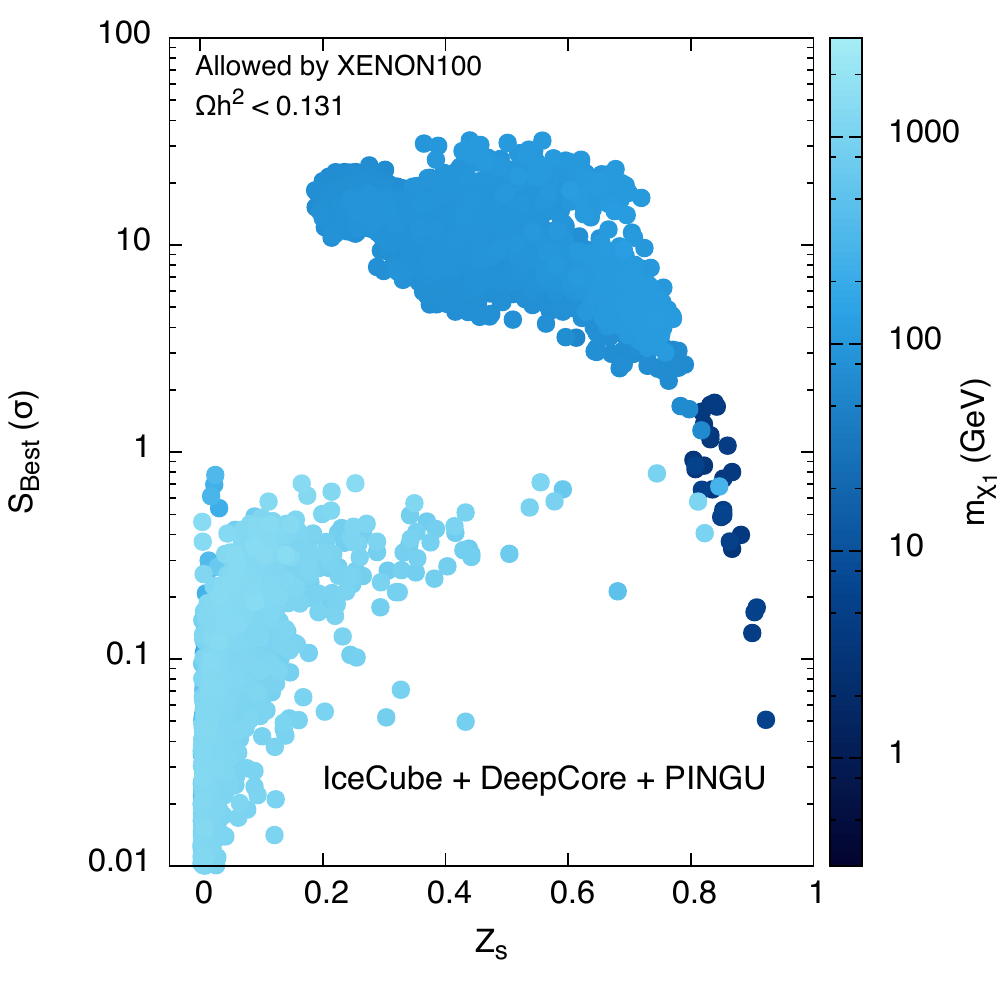}
\end{minipage}
\begin{minipage}{0.49\linewidth}
\includegraphics[width=\linewidth]{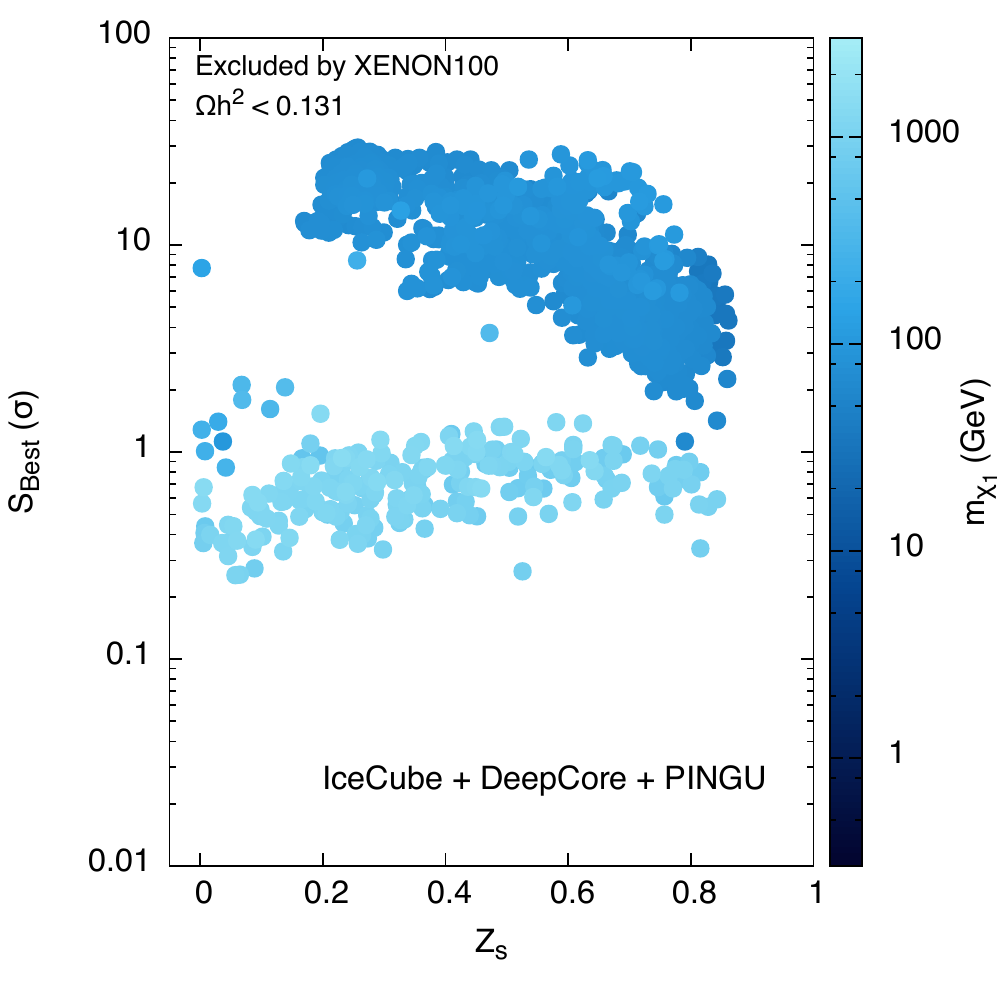}
\end{minipage}
\end{center}
\caption{
Best significance (from either of IceCube, DeepCore or PINGU) obtained for model points allowed (left) or excluded
(right) by XENON100, against the corresponding $Z_s$ and $m_{\neut{1}}$.}
\label{fig:Z_s-vs-signi}
\end{figure}

In figure~\ref{fig:Z_s-vs-signi} we show again the $S_\mathrm{Best}$,
but with the singlino fraction now on the $x$-axis and the DM mass given by
the colour code. The figure further illustrates that the IceCube
experiment can have a reasonable sensitivity to an abundance of points 
with a wide range of singlino fractions.

\begin{figure}[t]
\begin{center}
%\begin{minipage}{0.32\linewidth}
%\includegraphics[width=\linewidth]{m_neut1-years_BEST-PassXenon1T-DDAllowed}
%\end{minipage}
\begin{minipage}{0.49\linewidth}
\includegraphics[width=\linewidth]{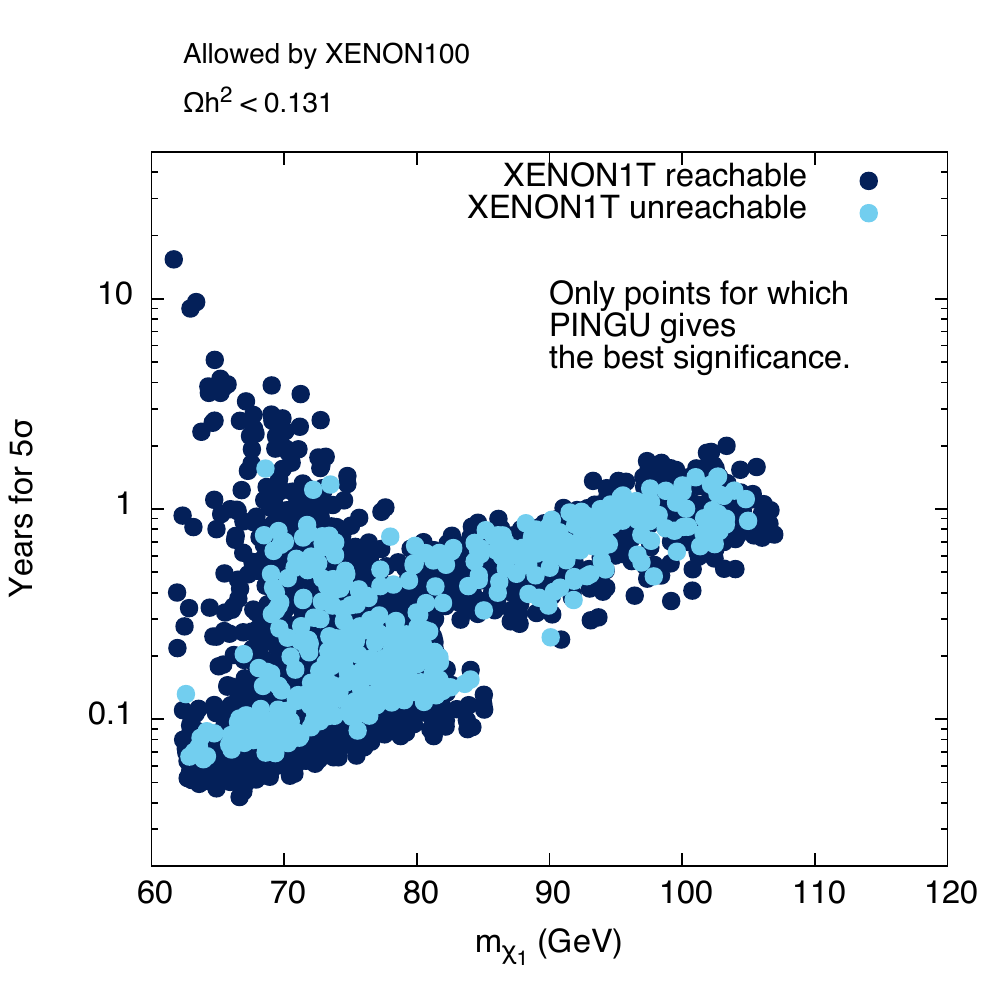}
\end{minipage}
\begin{minipage}{0.49\linewidth}
\includegraphics[width=\linewidth]{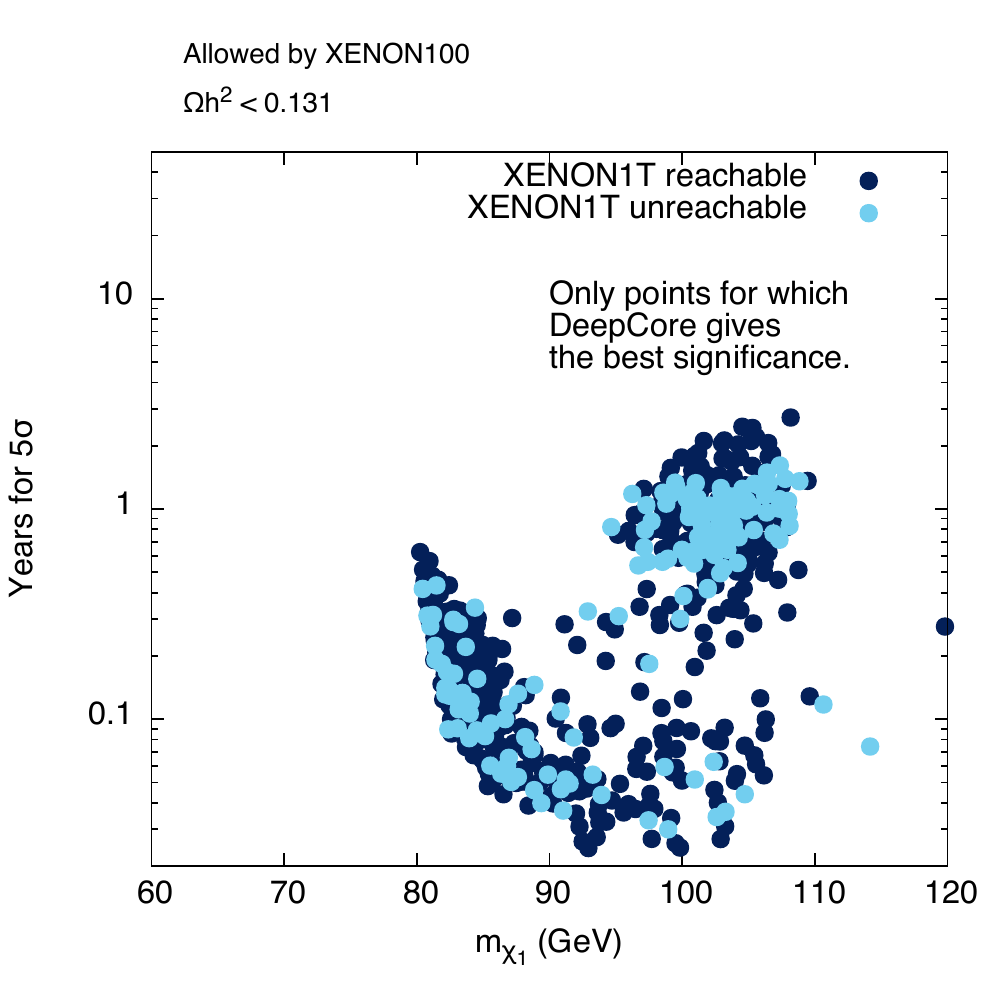}
\end{minipage}
\end{center}
\caption{
Number of calendar years required to reach $S=5\sigma$
as a function of the $\neut{1}$ mass.
The left plot shows only points where PINGU gives the best significance per calendar year,
whereas the right plot illustrates in a similar way the case of DeepCore.
Points not reachable by XENON1T with the projected sensitivity
are shown in a lighter shade on top of the remaining points.
We conclude that DeepCore and PINGU complement each other,
and can access model points that are out of reach of the projected XENON1T sensitivity.
The number of years has been extrapolated from the averaged 1-year significance. For PINGU we assume 
continuous data-taking over the calendar year.}
\label{fig:deepcore-vs-xenon1t}
\end{figure}

Finally, in figure~\ref{fig:deepcore-vs-xenon1t} we highlight the
complementarity of DeepCore/PINGU and the XENON direct detection experiment.
The figure shows the $m_{\neut{1}} \sim \unit[100]{GeV}$ model points
which can not be excluded
by the projected XENON1T 90\% CL limits in a light shade on top of
the ones that can be excluded, shown in darker points.
As seen earlier, a subset of the points is already excluded by the 
available DeepCore one-year data.
We note that more parameter space points, some of which will not
be probed by XENON1T, will be testable with each subsequent year of
data-taking by the DeepCore detector. 
The PINGU detector, when it enters operation, will also be able to 
test model points reachable neither by XENON1T nor by DeepCore due to 
the low $m_{\neut{1}}$, thus illustrating the complementarity of PINGU and 
DeepCore.

\section{Summary and conclusions\label{sec:summary}}

In this article we have analysed in depth the potential of the IceCube/DeepCore 
detectors and the proposed low energy extension, PINGU, to probe the higgsino-singlino 
sector of the NMSSM. We have presented our findings from the NMSSM parameter space scans,
which concentrated on model points predicting a
lightest neutralino with a non-vanishing singlino fraction, and hence 
a non-MSSM-like dark matter candidate. In these scans we required such 
points to survive the most important experimental constraints
from Higgs boson searches as well as from $b$-physics and relic
density measurements. 

We have used the $\nu_\mu$ spectra from DM annihilation in the Sun to test the model 
points of our interest. We have evaluated the signal on a
point-by-point basis in the NMSSM and 
taken into account the expected atmospheric background in the detectors considered.
We have then estimated the statistical significance that can be obtained for each 
model point by each detector and found that a subset of all the points obtained from
our scans have already been ruled out by the one-year data from the IceCube experiment, 
while more points will be accessible with each subsequent year of data-taking. 
We have found that PINGU will indeed have sensitivity to neutralino masses
of the order of $\unit[10]{GeV}$, although the detector should
accumulate several years of livetime to reach a 
significance of 2$\sigma$ for some of the points with a low $\neut{1}$ mass. 

We have also emphasised the complementarity of IceCube/DeepCore/PINGU searches to 
the ton-scale direct detection experiments, by highlighting parameter
space regions that may not be probed by the latter but will be accessible 
at the former.

%-------------------------------------------------------------------------------------------------
\acknowledgments
%-------------------------------------------------------------------------------------------------

We thank Matthias Danninger for providing the numerical values of the
effective areas that are shown in figure~7.17 of~\cite{DanningerThesis2011}.  
We thank the IceCube collaboration for providing their code NeutrinoFlux\ \cite{atmospheric} that we used to 
compute the atmospheric neutrino flux, and David Boersma, Olga Botner, Teresa Montaruli and Anne Schukraft for help with the code.
This work was funded in part by the Swedish Research Council under
contracts 2007-4071, 621-2011-5107 and 621-2011-5109. SM is
supported in part by the Korea Ministry of Science, ICT and Future Planning, 
Gyeongsangbuk-Do and Pohang City for Independent Junior Research 
Groups at the Asia Pacific Center for Theoretical Physics.
The numerical scans were in part performed on the computational
resources provided by the Swedish National Infrastructure
for Computing (SNIC) at Uppsala Multidisciplinary Center
for Advanced Computational Science (UPPMAX) under Projects p2013257 and SNIC 2014/1-5.

%-------------------------------------------------------------------------------------------------
\bibliographystyle{JHEP}
\bibliography{bibnmssm}
%-------------------------------------------------------------------------------------------------

\end{document}